\documentclass[twocolumn,floatfix,aps,prb]{revtex4}
\usepackage{graphicx}
\usepackage{amsmath}
\usepackage{amssymb}
\usepackage{bm}
\usepackage{color}

\newcommand{\sys}{\mathcal{S}}
\newcommand{\sil}{\mathcal{S}_{i,l}}
\newcommand{\dsil}{\partial\mathcal{S}_{i,l}}

\newcommand{\ham}{\hat{H}}
\newcommand{\Op}{\hat{O}}
\newcommand{\dOl}{\partial\hat{O}_{i,l}}

\newcommand{\Ol}{\hat{O}_l}
\newcommand{\Olm}{\hat{O}_{l-1}}
\newcommand{\oip}{\hat{o}_{i+1}}
\newcommand{\dtauil}{\partial\tau_{i,l}}
\newcommand{\dtauik}{\partial\tau_{i,k}}
\newcommand{\taui}{\tau_i}
\newcommand{\tauj}{\tau_j}

\newcommand{\tauil}{\tau_{i,l}}

\newcommand{\tauilm}{\tau_{i,l-1}}
\newcommand{\bracktauil}{t_{i,l}}
\newcommand{\bracktauilm}{t_{i,l-1}}

\newcommand{\minev}{\rho}
\newcommand{\trace}{\text{Tr}}
\newcommand{\aI}{a_{\Sigma}}

\newcommand{\veca}{\vec{a}}
\newcommand{\vecadag}{\vec{a}^\dag}
\newcommand{\locmat}{\mathcal{C}}

\newcommand{\numopsl}{\mathcal{N}_l}
\newcommand{\numopslp}{\mathcal{N}_{l+1}}
\newcommand{\Pp}{\mathcal{P}}
\newcommand{\Hh}{\mathcal{H}}
\usepackage{hyperref}
 \hypersetup{
     bookmarks=false,         
     unicode=false,          
     pdftoolbar=false,        
     pdfmenubar=true,        
     pdffitwindow=false,     
     pdfstartview={FitH},    
     pdftitle={Explicit construction of local conserved operators in disordered many-body systems},    
     pdfauthor={},     
     pdfsubject={},   
     pdfcreator={},   
     pdfproducer={}, 
     pdfkeywords={many-body localization} {disordered systems}, 
     pdfnewwindow=true,      
     colorlinks=true,       
     linkcolor=black,          
     citecolor=blue,        
     filecolor=magenta,      
     urlcolor=blue           
 }

\begin{document}

\title{Explicit construction of local conserved operators in disordered many-body systems}
\author{T. E. O'Brien,$^{1}$ Dmitry A. Abanin,$^{2}$ Guifre Vidal,$^{3}$ and Z. Papi\'c$^{4}$}
\affiliation{$^{1}$Instituut-Lorentz, Universiteit Leiden, P.O. Box 9506, 2300 RA Leiden, The Netherlands}
\affiliation{$^{2}$Department of Theoretical Physics, University of Geneva, 24 quai Ernest-Ansermet, 1211 Geneva, Switzerland}
\affiliation{$^{3}$Perimeter Institute for Theoretical Physics, Waterloo, Ontario N2L 2Y5, Canada}
\affiliation{$^{4}$School of Physics and Astronomy, University of Leeds, Leeds, LS2 9JT, United Kingdom}

\begin{abstract}
The presence and character of local integrals of motion -- quasi-local operators that commute with the Hamiltonian -- encode valuable information about the dynamics of a quantum system. In particular, strongly disordered many-body systems can generically avoid thermalisation when there are extensively many such operators. In this work, we explicitly construct local conserved operators in $1$D spin chains by directly minimising their commutator with the Hamiltonian. 
We demonstrate the existence of an extensively large set of local integrals of motion in the many-body localised phase of the disordered XXZ spin chain. These operators are shown to have exponentially decaying tails, in contrast to the ergodic phase where the decay is (at best) polynomial in the size of the subsystem.  We study the algebraic properties of 
localised operators, and confirm that in the many-body localised phase they are well-described by ``dressed" spin operators.
 \end{abstract}
\maketitle

\section{Introduction}\label{sec:intro}

Isolated quantum systems (i.e., those disconnected from any thermal bath) evolve according to unitary evolution. Recently there has been much interest in understanding and classifying the possible outcomes of such evolution for generic systems, e.g., those containing many particles, subject to local potentials and interactions, and in the presence of quenched disorder degrees of freedom. For simplicity, we consider systems defined on a lattice, and therefore with a local Hilbert space of finite dimension (e.g., two in the case of spin-1/2 qubits). The total Hilbert space, being a tensor product of local Hilbert spaces, is exponentially large in lattice size $L$, which makes the problem still very difficult to treat in general. 

One possible outcome that is relatively well understood is ergodic evolution: a system, prepared in an arbitrary initial state, evolves towards local thermal equilibrium at long times. 
This process is a result of the ``eigenstate thermalisation hypothesis" (ETH), which governs the underlying structure of individual (many-body) eigenvectors of ergodic systems~\cite{DeutschETH,SrednickiETH,RigolNature}. 
More recently, there has been a surge of activity focused on understanding a distinct class of systems which undergo \emph{non-ergodic} dynamics. A well-known example of such systems is the Anderson insulator~\cite{Anderson58}. Anderson localisation is a generic property of low-dimensional systems which is not sensitive to a particular type of lattice or disorder, but only applies to non-interacting systems. When interactions are comparable in strength to hopping and disorder energy scales, the system may exhibit \emph{many-body localisation}, a type of localisation in the many-body Hilbert space~\cite{Mirlin05,Basko06,OganesyanHuse}. 

The interest in many-body localised (MBL) systems is partly fundamental, as they provide a new paradigm for non-ergodic systems that violate the basic premise of equilibrium statistical mechanics. On the other hand, MBL systems have measurable properties that distinguish them from both Anderson insulators and ergodic systems. For example, their dynamics is governed by slow dephasing between different parts of the system~\cite{Znidaric08, Moore12, Serbyn13-1, Vosk13}. On the other hand, it was shown that quantum information can be recovered in MBL systems using spin-echo techniques~\cite{Serbyn_14_Deer}, and there are proposals that exotic types of order can be stabilised using the MBL mechanism~\cite{HuseSondhi,Bauer13,Bahri,Kjall14}, which may be applicable to designing quantum information processing schemes.

\begin{figure}
\includegraphics[width=\columnwidth]{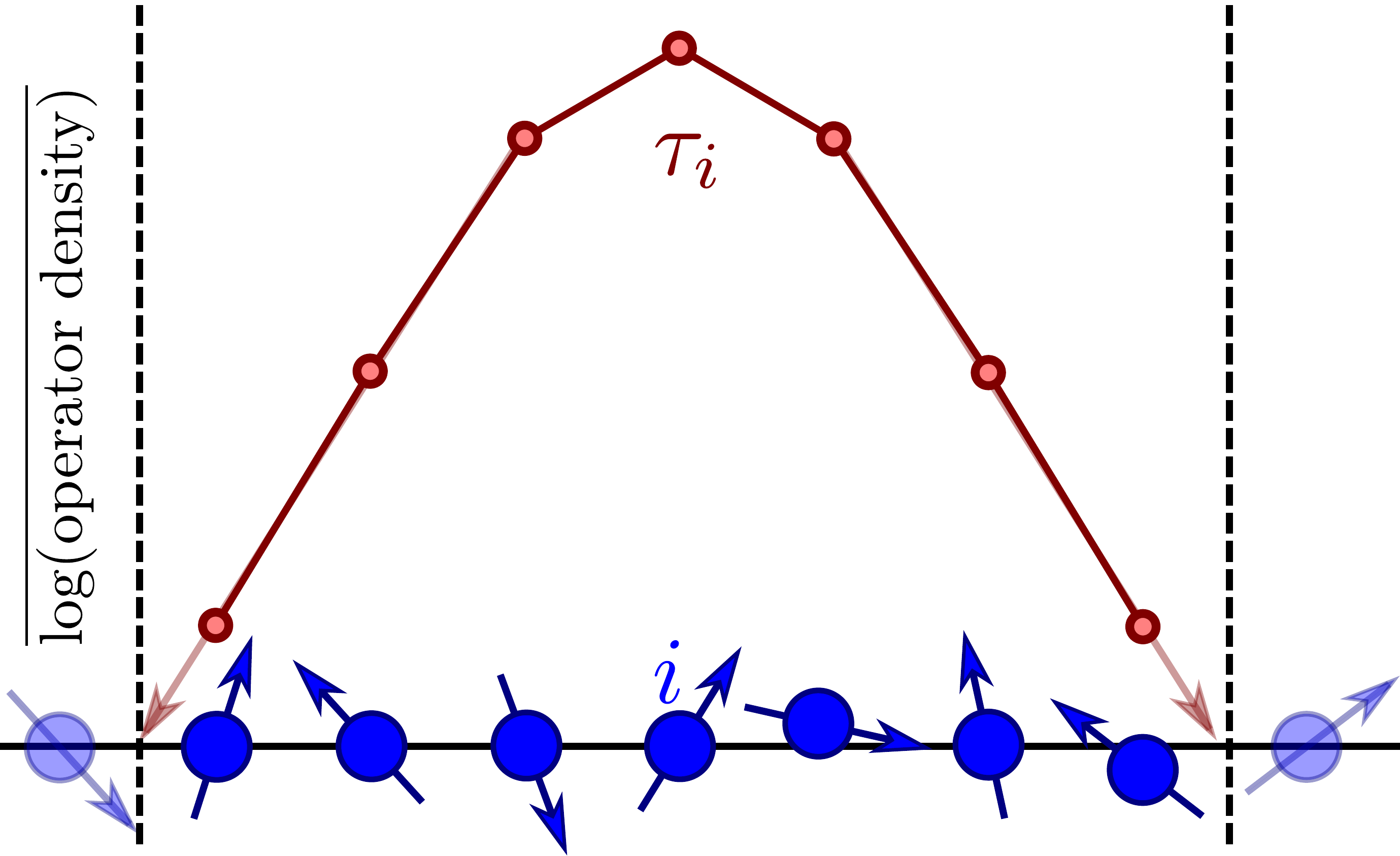}
\caption{\label{fig:schematic} (Color online) A typical integral of motion $\tau_i$ in the many-body localised phase of the disordered XXZ model.  $\tau_i$ decays exponentially away from the central site $i$. The operator $\tau_i$ is determined using the method described in Sec. \ref{sec:method}, and its spatial profile  is obtained by plotting the disorder-averaged logarithm of the operator density, Eq.~\eqref{opdensity}. The system is XXZ spin chain with interaction $V=0.5t$ and disorder $\Delta=5t$ in Eq. \eqref{XXZHam}.}
\end{figure} 

In order to explain the basic phenomenology of MBL systems, including  their failure to thermalise, a picture of \emph{Local Integrals of Motion} (LIOMs) has been put forward~\cite{Serbyn13-2,Huse13}. According to this picture, the basic mechanism of MBL is similar to integrable models \cite{Sutherland}: there emerges an extensive number of operators (``conserved charges") $\tau_i$, which commute amongst themselves $\left[  \tau_i, \tau_j \right] = 0$ as well as with the Hamiltonian, $\left[\hat H, \tau_i \right]=0$. A special property of MBL systems is that $\taui$ have eigenvalues $\pm 1$, thus they resemble the bare spin-$1/2$ operators, and generically there are $L$ such operators in a lattice system of size $L$. This means that any Hamiltonian eigenvector can be specified by the conserved quantum numbers corresponding to operators $\tau_i$, $i=1,\ldots, L$. Because of this extensive number of emergent quantum numbers (that by definition do not change during unitary evolution), the thermalisation of the system is prevented as the MBL state retains the memory of its initial condition. The difference between integrable models and MBL systems is in the form of individual $\tau_i$: in the integrable case, each $\tau_i$ is an extended sum of local operators, while in the MBL case each $\tau_i$ is a single local operator, up to corrections that vanish exponentially  with distance to the core. The subleading (exponentially suppressed) corrections are important, as they cause the distinction between Anderson and MBL insulators. For example, the presence of tails in LIOMs is responsible for the dephasing dynamics and the spreading of entanglement in MBL systems \cite{Moore12, Serbyn13-1}, which does not occur in Anderson insulators. 

The extensive number of LIOMs can therefore be viewed as a defining characteristic of MBL systems. For this reason, several previous works have explored ways of constructing the LIOMs for MBL systems. In Refs. \onlinecite{Chandran14, Kim14}, a local operator is  time-evolved with the Hamiltonian and averaged over time. 
Other works~\cite{Imbrie14,Ros14} have perturbatively constructed LIOMs
of the form written down in Refs.~\onlinecite{Serbyn13-1,Huse13}. In particular, Ref.~\onlinecite{Imbrie14} rigorously proves the existence of LIOMs near the classical limit of a 1D spin model. On the other hand,  Ref. \onlinecite{Ros14} follows the perturbative treatment laid out in Ref. \onlinecite{Basko06} near Anderson localised free fermions.  Most recently, in Ref. \onlinecite{Rademaker} an iterative construction for LIOMs was proposed based on a similarity transformation (see also Ref. \onlinecite{MonthusToda} and Ref. \onlinecite{He16}), while there have also been attempts at constructing the LIOMs via quantum Monte Carlo \cite{Inglis16}. 

In this work, we perform an explicit construction of local conserved operators, focusing on the many-body localised phase in $1$D spin chains. We follow the method introduced in Ref. \onlinecite{KimCirac} where a slow-moving mode was identified in ergodic Hamiltonian and Floquet systems. 
Our method directly minimises the commutator, with the system's Hamiltonian, of an operator restricted to a small subregion of an infinite chain.
The main result is an unambiguous signal of the exponentially decaying tails of localised operators in the MBL phase of the disordered XXZ spin chain. This is in contrast to the polynomial scaling of the analogous operator in the ergodic phase. Further, we demonstrate that an \emph{extensively large set} of localised operators can be found in the MBL phase, in contrast to the ergodic phase. We also study the algebraic properties of localised operators, finding that when the system is localised they resemble spin operators with exponentially decaying tails.

As a consequence, our results allow to directly visualise the LIOMs that continue to play a vital role in the theoretical studies of MBL systems, see. Fig. \ref{fig:schematic}. By plotting the disorder-averaged logarithm of the operator density, defined in Eq.~\eqref{opdensity} below, we map out the spatial profile of a typical integral of motion $\tau_i$ in the MBL phase of the disordered XXZ model.  This indeed confirms that $\tau_i$ decays exponentially in the distance from the site $i$. 

The remainder of this paper is organised as follows. In Sec. \ref{sec:def} we introduce the notation and definitions. In Sec. \ref{sec:model} we briefly review the properties of the disordered XXZ chain -- the prototype model of the MBL phase and the localisation-delocalisation transition. In Sec. \ref{sec:method} we present our method for explicitly constructing operators supported on a finite region which minimise the commutator with the given Hamiltonian. The method can be formulated as finding the approximate zero modes of a certain matrix, which we call the ``commutant matrix". In Sec. \ref{sec:decay} we study the scaling of the smallest eigenvalue of the commutant matrix with the size of the subsystem. We confirm the polynomial decay of this eigenvalue in the ergodic phase as previously found in Ref. \onlinecite{KimCirac}, and show that it decays exponentially in the MBL phase. In Sec. \ref{sec:counting} we show that the difference between ergodic, integrable and MBL phases can also be characterised by the \emph{counting} of approximate zero modes in the commutant matrix. In particular, the MBL phase is distinguished by having an extensively large manifold of such zero modes in contrast to other phases. In Secs. \ref{sec:characteristics} and \ref{sec:corr} we examine in more detail the properties of the constructed LIOMs, such as the ``operator participation ratio" and their spectral measures. Our conclusions are presented in Sec. \ref{sec:conc}, where we also discuss some future directions.

\section{Background and definitions}\label{sec:def}

In the following, for simplicity, we restrict ourselves to one-dimensional spin chains with local Hamiltonians (sums of terms with local support). Our methods can be directly extended to higher dimensional systems, although this comes with considerable  computational cost.

Consider a spin chain $\sys$. Around any spin $i$ in a chain $\sys$, a series of concentric one-dimensional spheres of radius $l$ (connected line segments and their endpoints) can be formed:
\begin{align}
\sil&=\{i-l,i-l+1,\ldots,i+l-1,i+l\},
\end{align}
where the boundary of this segment is denoted by
\begin{equation}
\dsil = \{i-l,i+l\}.
\end{equation}
We are interested in the behaviour of operators within the areas $\sil$. We can focus on traceless operators $\Op$ satisfying  $\trace[\Op]=0$, as any operator can be written as a basis-independent linear combination of a traceless part and the identity (which has no local character).

A traceless operator $\Op$ acting on the Hilbert space of $\sys$ ($\Hh_{\sys}$) may then be decomposed into a limit of operators $\Op=\lim_{l\rightarrow\infty}\Ol$:
\begin{equation}
\Ol=\trace_{~\overline{\sil}}~\Op,
\end{equation}
where the trace is taken over the complement of $\sys_{i,l}$. We write the difference $\Ol-\Olm=:\dOl$. We say $\Op$ is \emph{centred at $i$ with exponential tails} if there exists an exponential function $A\exp(-l/\xi)$ that bounds $||\dOl||$. Here, $\xi$ is a characteristic length scale that that governs some of the correlations involving $\Op$. It is tempting to identify $\xi$ as the ``many-body localisation length'', which should diverge at the MBL transition. Note, however, that the decay of correlations within the MBL phase could be governed by multiple lengths \cite{Huse13}. Some of those may diverge at the phase transition, while others may not, e.g., the length scale $\xi_c$ defined below, which controls the spreading of entanglement, is expected to remain finite at the phase transition. Our analytic arguments below are restricted to systems deep inside the MBL phase, where $\xi$ is finite.

Given a Hamiltonian $\ham$ acting on $\sys$, an \emph{integral of motion} (IOM) is an operator $\Op$ that commutes with $\ham$.
In contrast to this generic operator, the MBL phase \cite{Serbyn13-2,Huse13} supports extensively many integrals of motion, each centred around some site $i$ in  $\sys$, with exponentially decaying tails. We call such an operator a \emph{local integral of motion} (LIOM) and denote it by $\taui$. The exponential decay of $\taui$ in the MBL phase is governed by some length scale $\xi$, which is related to the many-body localisation length. In Sec. \ref{sec:method} below, we will outline the procedure how such $\tau_i$ can be directly constructed. The result of such a calculation gives a typical $\tau_i$ illustrated in Fig. \ref{fig:schematic}. In this figure, we plot the logarithm of the operator density (defined in Eq. \eqref{opdensity} below). The important sign of the MBL phase is the straight line fit to the tails on either side, as this corresponds to exponential decay.

As $\ham$ is a sum of terms with local support, its commutator with the restricted $\tauil$ must live in the area around $\dsil$ and be cancelled by operators that extend to this area. Thus, we can bound it as:
\begin{equation}
||[\ham,\tauil]||\leq C||\dtauil||\leq C'\exp(-l/\xi).
\label{eq:LIOM_bound}
\end{equation}
We intend to approximate $\tauil$ in each $\sil$ by finding the \emph{approximate local integral of motion} (ALIOM) $\bracktauil$ in $\sil$ that minimises the commutator with $\ham$. 

\section{Model: the disordered XXZ spin chain}\label{sec:model}

A prototype model of a system with an MBL phase is the disordered XXZ spin chain \cite{PalHuse}. This consists of a one-dimensional spin-$1/2$ chain with the Hamiltonian
\begin{eqnarray}
\nonumber \ham&=&\frac{t}{4}\sum_{j=1}^{N-1}\left(\sigma_j^x\sigma_{j+1}^x+\sigma_j^y\sigma_{j+1}^y\right)\\&+&\frac{V}{4}\sum_{j=1}^{N-1}\sigma_j^z\sigma_{j+1}^z +\frac{1}{2}\sum_{j=1}^Nh_j\sigma_j^z, \label{XXZHam}
\end{eqnarray}
with $\sigma_j^a$, $a=\{0,x,y,z\}$ being the Pauli operator acting on site $j$ and satisfying $(\sigma_j^a)^2=1$, and $h_j$ selected randomly from a uniform distribution over $[-\Delta,\Delta]$. We consider open boundary conditions. The model has two independent parameters ($V$ and $\Delta$), assuming the overall energy scale is fixed by $t$. Via the Jordan-Wigner transformation, this system is equivalent to the (spinless) $1d$ Hubbard model, where $\Delta$ is random on-site disorder. 

The disordered XXZ model was found to capture several interesting physical regimes. Let us assume $V=t$. At zero disorder $\Delta=0$, the model is integrable and thus fails to thermalise. For disorder $\Delta<\Delta_c$ with $\Delta_c\approx 3.5-4$ the critical value of disorder, the system is ergodic, in the sense that nearly all eigenstates obey the ETH \cite{Kim_ETH} For small values of $\Delta\lesssim 0.55$, entanglement spreads ballistically \cite{KimBallistic} and obeys the ``volume law" in system's eigenstates \cite{Znidaric16}. This means that the von Neumann entropy $S_{ent}$, evaluated for a system bipartitioned into $A$ and $B$ subsystems (such that $L_A+L_B=L$), scales proportionally to the number of degrees of freedom in $A$: $S_{ent}\propto L_A$. This is a typical property of ergodic systems, and a similar relation holds even for completely random vectors in the Hilbert space. At slightly larger disorder $0.55\lesssim\Delta\lesssim \Delta_c$, the system exhibits sub-ballistic entanglement spreading, and sub-diffusive charge transport \cite{Znidaric16}.

Studies of the inverse participation ratio (IPR) and eigenvalue statistics \cite{PalHuse} have demonstrated the existence of a finite-$V$ phase transition in the range  $3 \lesssim \Delta \lesssim 4$. There is evidence that this transition may be complicated by the presence of a mobility edge \cite{Alet14, Serbyn15}, making the position of the phase transition dependent on energy density, but it has been suggested
that this may be due to finite-size effects \cite{SchiulazMobility}. 

Finally, at strong disorder $\Delta \gtrsim 4$, the system enters the MBL phase where nearly all eigenstates are localised (apart from rare resonant states). In the MBL phase, the entropy of most eigenstates scales as the area of the subsystem, i.e., $S_{ent} \propto const$ in 1D. This means, on average, there is far less entanglement in MBL states compared to ergodic systems. This ``area law" for entropy \cite{Huse13, Bauer13, Friesdorf} is similar to what is found in the ground states of gapped systems \cite{Plenio}. The special property of MBL phase is that the area law holds throughout the spectrum, even at arbitrarily high excited eigenstates. This also has consequences for the dynamics in the MBL phase, for example during the global quench from an initial product state, the entanglement spreads logarithmically in time \cite{Znidaric08, Moore12, Serbyn13-1}: $S_{ent}(t) \propto \xi_c \ln(Vt/\hbar)$, where $\xi_c$ is an effective length controlling the spread of entanglement(note that $\xi_c$ is expected to remain finite at the MBL transition \cite{Huse13}).

As mentioned in Sec. \ref{sec:intro}, the fundamental reason why entanglement and dynamics are so constrained in the MBL phase compared to the ergodic phase, is the appearance of extensively many LIOMs. In order to fully understand the properties of MBL systems, it is therefore desirable to have a method that has direct access to such operators. In the following Section, we describe our method for constructing such operators directly by solving an eigenvalue problem.

\section{Method}\label{sec:method}

We now introduce the method for explicitly constructing operators that ``best commute" with the Hamiltonian, such as the one in Eq.(\ref{XXZHam}) (see also Ref. \onlinecite{KimCirac}). The method is general and can be applied to any model. In subsequent Sections, we will apply the method to construct local operators and study their scaling with system size in the MBL and ergodic phases of the disordered XXZ model. 

Given a spin chain $\sys$, we wish to minimise the commutator \cite{KimCirac}
\begin{equation}
\minev=\frac{\trace \left( \left[\ham,\Op\right]^\dag\left[\ham,\Op\right]\right)}{\trace\left( \Op^{\dag}\Op \right)},\label{eqn:minev}
\end{equation}
for operators $\Op$ that are sums of terms with support on a subsystem $S$ of $\sys$. We exclude the identity from this sum, as it trivially commutes with $\ham$. This requirement is equivalent to requiring $\Op$ be traceless.

We now expand $\Op$ as a linear combination of operators from a basis for the set of traceless operators with support in $S$. A natural basis to take (for spin-$1/2$ systems) is the set $\Pp_S$ of local products of Pauli operators. Each operator in $\Pp_S$ is defined by choosing a Pauli operator from $\{\sigma_i^0,\sigma_i^x,\sigma_i^y,\sigma_i^z\}$ for each site $i$, and taking the tensor product of the combination. $\Pp_S$ contains all possible such operators excluding the identity (which corresponds to the choice of $\sigma^0$ on all sites). Note that this gives $4^{|S|}-1$ elements in $\Pp_S$, making it a complete basis for operators with support on $S$ when combined with the identity.

With this choice we write
\begin{equation}
\Op=\sum_{\Sigma\in\Pp_S} \aI\Sigma.
\end{equation}
If we substitute this into \eqref{eqn:minev}, and write $\veca$ as the vector of individual $\aI$ terms, we can rewrite (using the fact that operators in $\Pp_S$ satisfy $\trace\left(\Sigma_i^{\dag}\Sigma_j\right)=\delta_{i,j}4^{|\sys|}$)
\begin{equation}
\minev=\frac{\vecadag\locmat\veca}{\vecadag\veca},\label{eqn:rhoeq}
\end{equation}
where we introduced a ``commutant matrix" $\locmat$, defined as
\begin{equation}\label{eq:commutant}
\locmat_{\Sigma_i,\Sigma_j}=\trace ([\ham,\Sigma_i]^\dag [\ham,\Sigma_j])4^{-|\sys|}.
\end{equation}
In order to minimise $\minev$, we need $\frac{\partial\rho}{\partial\vecadag}=0$. Multiplying both sides of Eq. \eqref{eqn:rhoeq} by $\vecadag\veca$, and differentiating by $\vecadag$ gives
\begin{equation}
\rho\veca=\locmat\veca
\end{equation}
Thus, the problem reduces to finding the eigenvector $\vec{v}$ of the commutant matrix $\locmat$ corresponding to the lowest eigenvalue $\minev$. From this we can construct the \emph{best-localised} operator $\Op_{\minev}=\sum_{\Sigma\in\Pp_{S}}\vec{v}_{\Sigma}\Sigma$.

Note that $\locmat$ is positive semi-definite, so all eigenvalues are $\geq 0$. To show this, let us define $S+\ham$ as the set of sites connected to $S$ by terms in $\ham$ (i.e. site $r$ lies in $S+\ham$ if and only if there is some site $s$ in $S$ and some $\Sigma_{rs}\in\Pp_{\sys}$ that contributes to $\ham$ and acts non-trivially on both sites $r$ and $s$). Then, if $\Sigma_i\in\Pp_S$, $[\ham,\Sigma_i]$ acts non-trivially only on $S+\ham$, and we can write 
\begin{equation}
\left[\ham,\Sigma_i\right]=\sum_{\Omega\in\Pp_{S+\ham}}c_{\Omega,i}\Omega,\;\;\; c_{\Omega,i}=\trace\left(\Omega^{\dag}[\ham,\Sigma_i]\right)4^{-|\sys|}.
\end{equation} 
Then, if $\Sigma_j\in\Pp_S$ also, we have 
\begin{equation}
\trace([\ham,\Sigma_j]^{\dag}[\ham,\Sigma_i])4^{-|\sys|}=\sum_{\Omega}c_{\Omega,j}^{\dag}c_{\Omega,i}.
\end{equation}
Then, from the definition of $\locmat$, we can immediately write $\locmat=C^{\dag}C$ with
\begin{equation}
C_{\Omega,\Sigma_i}=c_{\Omega,i}=\trace\left(\Omega^{\dag}[\ham,\Sigma_i]\right)4^{-|\sys|}.
\label{sd_eqn}
\end{equation}

This result implies that $\locmat$ does not have any eigenvalues other than what we are searching for; $\minev$ does indeed correspond to the eigenvalue closest to $0$. Furthermore, it is computationally far easier to construct the matrix $C$ and then calculate $\locmat=C^{\dag}C$ than directly calculate $\locmat$ from its definition.

\begin{figure*}
\begin{tabular}{cc}
\includegraphics[width=0.5\textwidth]{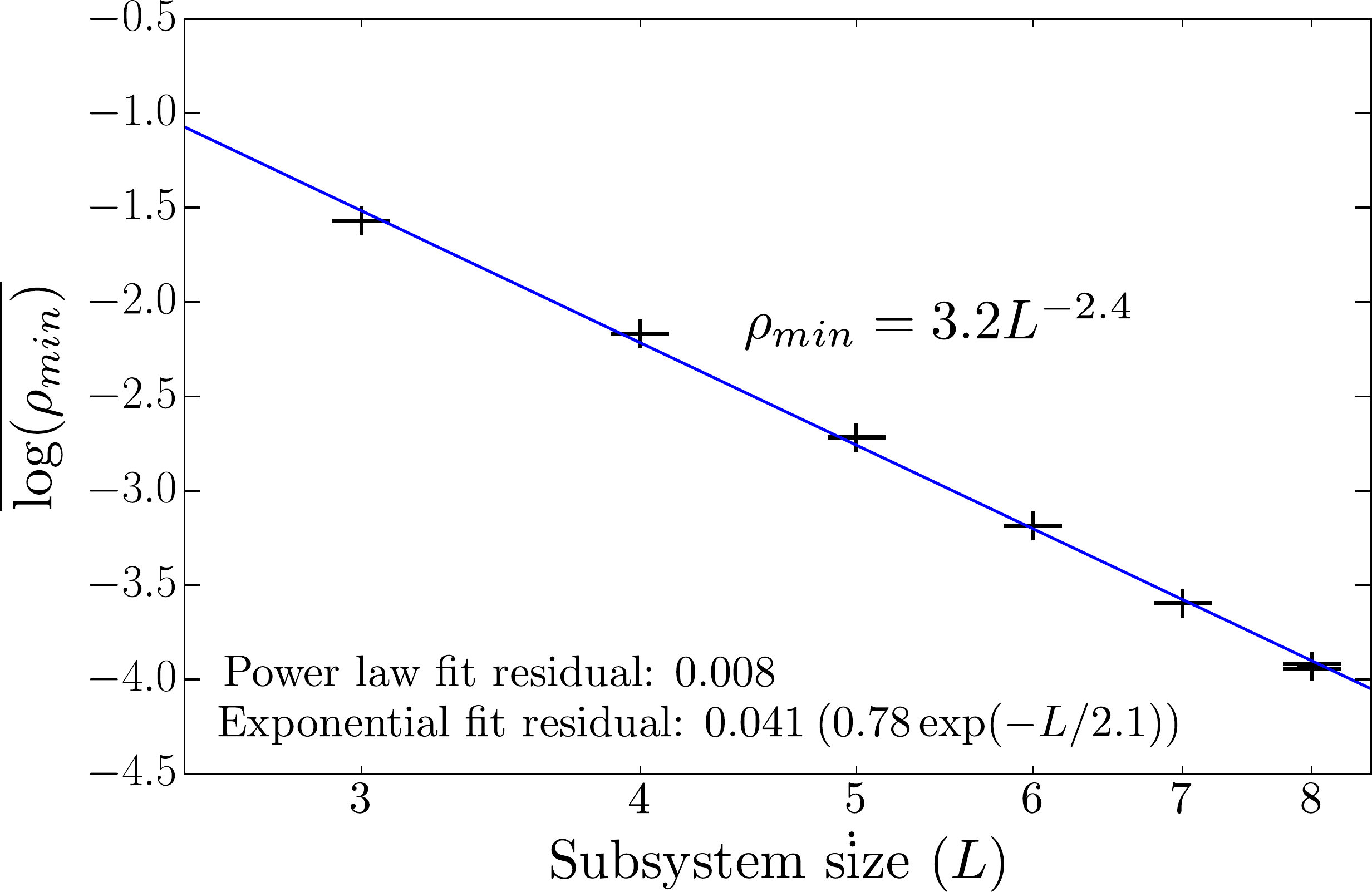}&
\includegraphics[width=0.5\textwidth]{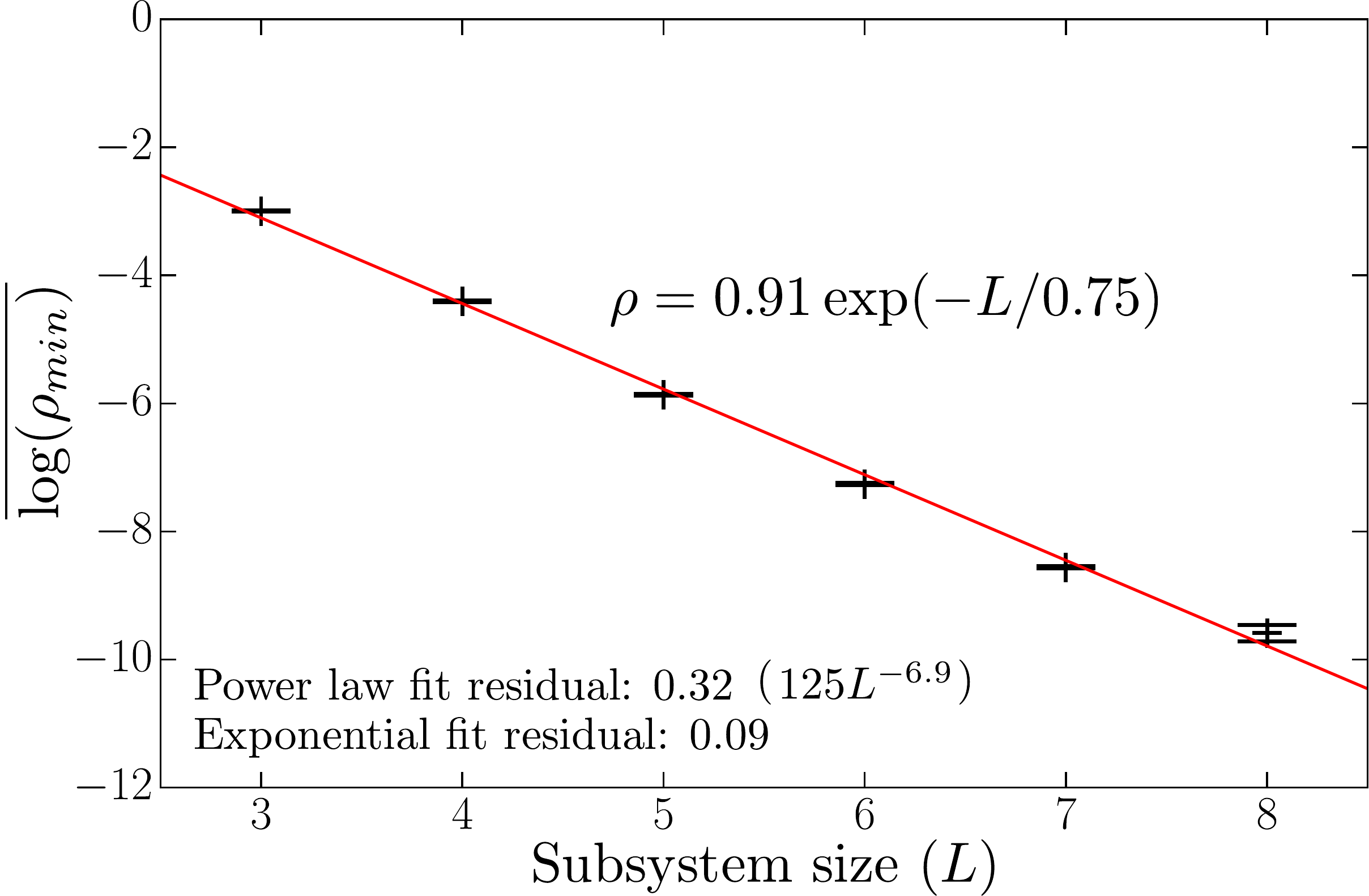}
\end{tabular}
\caption{\label{decay_fig} (Color online) Minimum eigenvalue of the commutant matrix for the disordered XXZ spin chain (\ref{XXZHam}) as a function of the subsystem size $L$. Parameters are $V=0.5t$, and $\Delta=0.5t$ (left) or $\Delta=5t$ (right). Points are averaged over either $10000$ disorder samples ($L=3,4,5,6,7$) or $200$ ($L=8$). Lines are polynomial (left) and exponential (right) fits to the points. Fit parameters are given in the figures.}
\end{figure*}

We briefly describe here the method used to construct Fig. \ref{fig:schematic}. Given a fixed subsystem $\sys$ of $7$ sites, we find the \emph{best-localised} operator $\Op_{\minev}$ that corresponds to the lowest eigenvalue of the localisation matrix. For each site $i\in\sys$, we calculate
\begin{equation}
1-||\trace_{\sys/i}\Op_{\minev}||^2\label{opdensity},
\end{equation}
where $\sys/i$ is the set of points with $i$ excluded. This is equivalent to finding the trace norm of the piece of $\Op_{\minev}$ with nontrivial support on $i$. This is then averaged for each site over $10000$ disorder realisations.

\section{Eigenvalue decay in the disordered XXZ spin chain} \label{sec:decay}

To begin our study of the disordered XXZ spin chain via the method explained in Sec. \ref{sec:method}, we demonstrate the difference in operator decay between the MBL and ergodic phases. In appendix \ref{app:decay_equivalence} we argue that this corresponds to the exponential decay of the lowest eigenvalue of the commutant matrix. 

We fix a subsystem $\sys_L$ of $L$ spins (labelled $i=1,\ldots,L$), and take all combinations of Pauli matrices acting on these spins as a basis of operators from which we construct the commutant matrix in Eq. (\ref{eq:commutant}). This matrix is then independent of Hamiltonian terms that do not act on these spins, and so for our XXZ model we need only embed this in a larger chain of $L+2$ spins. It is then possible to obtain the lowest eigenvalue of the commutant matrix for $L=8$ via iterative diagonalisation, e.g., using Arpack. This method is notably different in spirit from the diagonalisation of the Hamiltonian,
as the total size of the system can be assumed infinite and the boundary conditions play no role. 

In Fig. \ref{decay_fig}, we plot the disorder-averaged lowest eigenvalue of the commutant matrix for the range $L=3$ to $L=8$, in the ergodic ($V/t=0.5,\Delta/t=0.5$) and MBL ($V/t=0.5,\Delta/t=5$) regimes. The data shows \emph{typical} lowest eigenvalues of the commutant matrix (i.e., we average the logarithm of the lowest eigenvalue). Similar results are obtained when averaging the eigenvalue itself. In both cases, we attempt linear fits of $\ln(y)=ax+b$ (exponential decay) and $\ln(y)=a\ln(x)+b$ (power law decay). Found values of $a$ and $b$ are given in the figure. In the ergodic phase, the power law fit has a three times smaller residual than the exponential, whilst in the MBL phase the reverse occurs. Beyond the difference in the fit discrepancies, the power law fit to the results of the MBL system has almost $L^{-7}$ decay, which appears implausible. The power-law decay in the ergodic phase ($L^{-2.4}$) matches similar trends in other ergodic systems (Ref.~\onlinecite{KimCirac} finds a $L^{-2.6}$ decay in a different spin chain within an ergodic phase).

\section{Eigenvalue counting in the commutant matrix}\label{sec:counting}

In an ergodic system, Ref. \onlinecite{KimCirac} found a single slow operator with an eigenvalue that scales as a power law function of localisation area $L$. By contrast, the MBL phase is distinguished by the presence of not one but extensively many local integrals of motion. This suggests that one should not only focus on the ``ground state" of the commutant matrix, but the entire low-eigenvalue manifold of eigenvectors. In this Section, we analyse the structure of this manifold and provide counting rules for the complete set of localised operators than span this subspace.

\begin{figure}[hhh]
\includegraphics[width=\columnwidth]{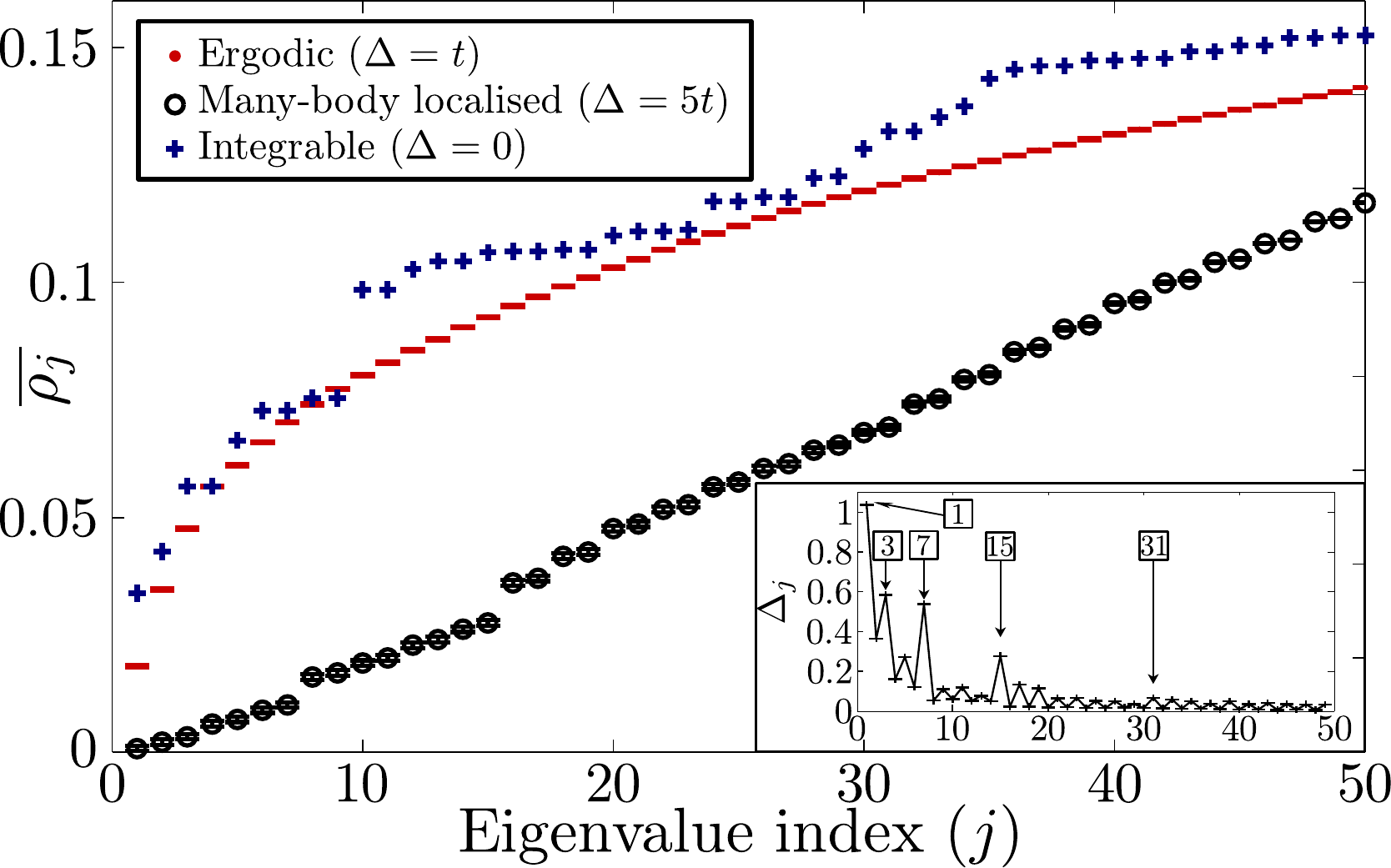}
\caption{\label{eigenspectrum} (Color online) The lowest 50 eigenvalues of the commutant matrix for systems in the integrable (red), ergodic (blue) and MBL (black) phases, with a subsystem of 7 sites. The results for the integrable system are exact, those for the ergodic and MBL systems are averaged over 5000 disorder realisations (error bars are shown in figure). (inset) Relative difference between the eigenvalues of the commutant matrix in the MBL phase, Eq. \eqref{eq:reldiff}. Points that correspond to changes between the distance $d_I$ in Fig. \ref{ALIOMFigure} have been marked, and a local maximum is visible at each point.}
\end{figure}

In Fig. \ref{eigenspectrum} we plot the averaged low-lying spectrum of the commutant matrix for a subsystem of length $L=7$, in the MBL ($\Delta=8t$), ergodic ($\Delta=1t$) and integrable ($\Delta=0$) regimes (all with $V=0.5t$). We see that at the lowest eigenvalues, the trend of the points in the MBL phase is slightly concave up; a large number of eigenvalues are clustered near zero. By comparison, the ergodic and integrable spectra are strongly concave down, and the lowest eigenvalue is atypically small.

Of interest is the clear presence of gaps in the MBL phase, which are not present when the system is ergodic. In order to display these more clearly, in the inset to Fig.  \ref{eigenspectrum} we plot the relative difference
\begin{equation}\label{eq:reldiff}
\Delta_j=2\frac{\rho_{j+1}-\rho_j}{\rho_{j+1}+\rho_j}
\end{equation}
(the specific labelling of the peaks in this plot is explained below). Interestingly, gaps are also present in the integrable limit, where new conserved operators have recently been constructed \cite{Prosen2011, Pereira2014, Ilievski2015}.  Pairs of degenerate eigenvalues in the integrable system are due to the $\sigma_x\leftrightarrow\sigma_y$ symmetry in the commutant matrix (these operators exist in separated blocks of the commutant matrix that are coupled by the disorder terms $\sigma_z$).

To explain the presence of gaps in the MBL case, let us consider the $L$ $\taui$ operators present in an $L$-site subsystem $\sys$ in the MBL phase. If we select a subset of sites $S\subset\sys$, the product $P_S=\prod_{i\in S}\taui$ will have exponentially decaying tails, as we can bound
\begin{equation}
||\partial P_{S,l}||<|S|A\exp(-l/\xi),
\end{equation}
with $A\exp(-l/\xi)$ the bounding function for the exponential decay of the LIOMs (which is system-dependent), and $l$ defined by the distance from the edge of $S$. This gives a total of of $2^L-1$ LIOMs centred in $\sys$. Each of these LIOMs should be approximated by a different eigenvector of the commutant matrix (as they satisfy $\trace(P_S^{\dag}P_{S'})=\delta_{S,S'}4^{|\sys|}$). If the approximation is good, the size of the corresponding eigenvalue will be proportional to the size of the tail of the LIOM that lies outside $\sys$, as per Eq. \eqref{eq:LIOM_bound}. The size of this tail grows exponentially with the minimum distance $d_S$ between any $i\in S$ and the boundary. Thus, we would expect groups of eigenvalues corresponding to $P_S$ operators with different $d_S$.

On the small system sizes studied, the situation is slightly complicated by disorder, which breaks the reflection symmetry within $\sys$, causing one side to be 'favoured' for decay over the other. To understand this, let us cut $\sys$ into left ($\sys_L$) and right ($\sys_R$) sides. A fixed disorder realisation $\{h_i\}$ will be stronger on one of the two sides; this can be roughly characterised by the standard deviation of $\{h_i,i\in\sys_a\}$ for $a=L,R$. Let us assume, without loss of generality, that the disorder on $\sys_L$ is stronger than that on $\sys_R$. Then, LIOM tails will have decay of the form $A_L\exp(-l/\xi_L)$ on the left but $A_R\exp(-l/\xi_R)$ on the right, with $\xi_R>\xi_L$. This implies that the operators $P_S$ that achieve $d_S$ on the right will have larger tails ($\propto\exp(-l/\xi_R)$) outside the boundary than operators $P_{S'}$ that achieve $d_{S'}$ on the left but \emph{not} the right (with tail size $\propto\exp(-l/\xi_L)$).

\begin{figure}[ht!]
\includegraphics[width=\columnwidth]{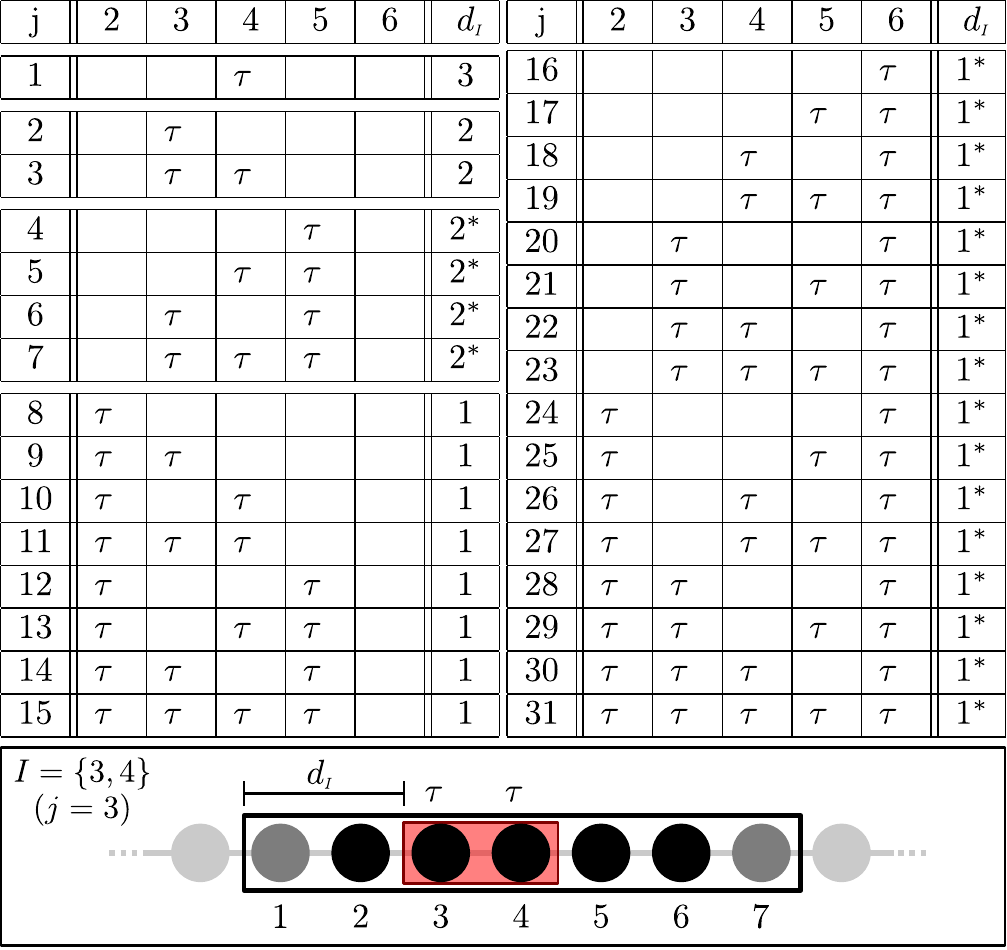}
\caption{\label{ALIOMFigure} (Color online) (top) Table of some combinations of local integrals of motion $P_S$ in a subsystem of width $7$ in a spin chain. On each row the single $\tau$ operators in the product $\prod_{i\in S}\tau_i$ are indicated, and the distance to the boundary $d_S$ is written on the right column. An asterisk denotes when the distance $d_S$ is to the boundary of slower decay (see text). (bottom) Example of a $P_S$ operator; in this case we have $P_S=\tau_3\tau_4$. If $P_S$ were approximated by our method inside the boxed region, the closest distance to the boundary is $d_S=2$ on the left, and we would expect the average $||[[P_S],\ham]||\propto\exp(-2/\xi)$, with $\xi$ the localisation length.}
\end{figure}

Continuing our assumption of stronger decay on the left, in Fig. \ref{ALIOMFigure} we order the $2^5-1=31$ $P_S$ operators by their distances $d_L$ and $d_R$. We also include a schematic of the scenario described above. Unless $\xi_R \gg \xi_L$, we would expect $\exp(-l/\xi_L)>\exp(-(l+1)/\xi_R)$, and so the tail size should be determined firstly by $\min(d_L,d_R)$ and secondly by whether $\min(d_L,d_R)=d_R$. We see that this gives groups corresponding exactly to the gaps seen in the MBL phase in Fig. \ref{eigenspectrum}. This shows that our method can approximate not just one, but several LIOMs.

The argument above does not explain why the peak height drops in Fig. \ref{eigenspectrum} (inset). We attribute this to interference from operators in the system other than the LIOMs, which may yet have a slow power-law decay. As the system size increases, we would expect these differences to grow until some saturation point, which would depend on the localisation length $\xi$. We would also expect the symmetry-breaking effect to disappear. Unfortunately we have not been able to study this due to our restrictions on subsystem size.

\section{ALIOM operator characteristics}\label{sec:characteristics}
 
If $||\trace_{\sil}\taui||/||\taui||\approx 1$, it is reasonable to expect the ALIOM $\bracktauil$ to be a good approximation to $\taui$. We now investigate whether $\bracktauil$ has the characteristics of a local spin.

At large disorder strength, the tails of $\taui$ should sit well within a large subsystem, and so $\bracktauil$ should itself be a sum of a small number of Pauli operators. We measure this by a means similar to the traditional inverse participation ratio for states: with again $\bracktauil=\sum_Ia_I\Sigma_I$ we can write
\begin{equation}
R(\bracktauil)=\sum_I|a_I|^4.
\label{IPRDef}
\end{equation}
This, or rather $R^{-1}$, can be thought of as a measure of how many of the Pauli operators contribute to $\bracktauil$; if we have $n$ equally contributing operators $\bracktauil=\frac{1}{\sqrt{n}}\sum_{j=1}^n\sigma_j$, $R^{-1}(\bracktauil)=n$. In general with uneven contributions $R^{-1}(\bracktauil)$ is not an integer, but it is bounded below by $1$. We dub $R^{-1}(\bracktauil)$ the \emph{operator participation ratio} (OPR). Note that the value of $R$ is dependent on the basis chosen for the commutant matrix.

\begin{figure}
\includegraphics[width=\columnwidth]{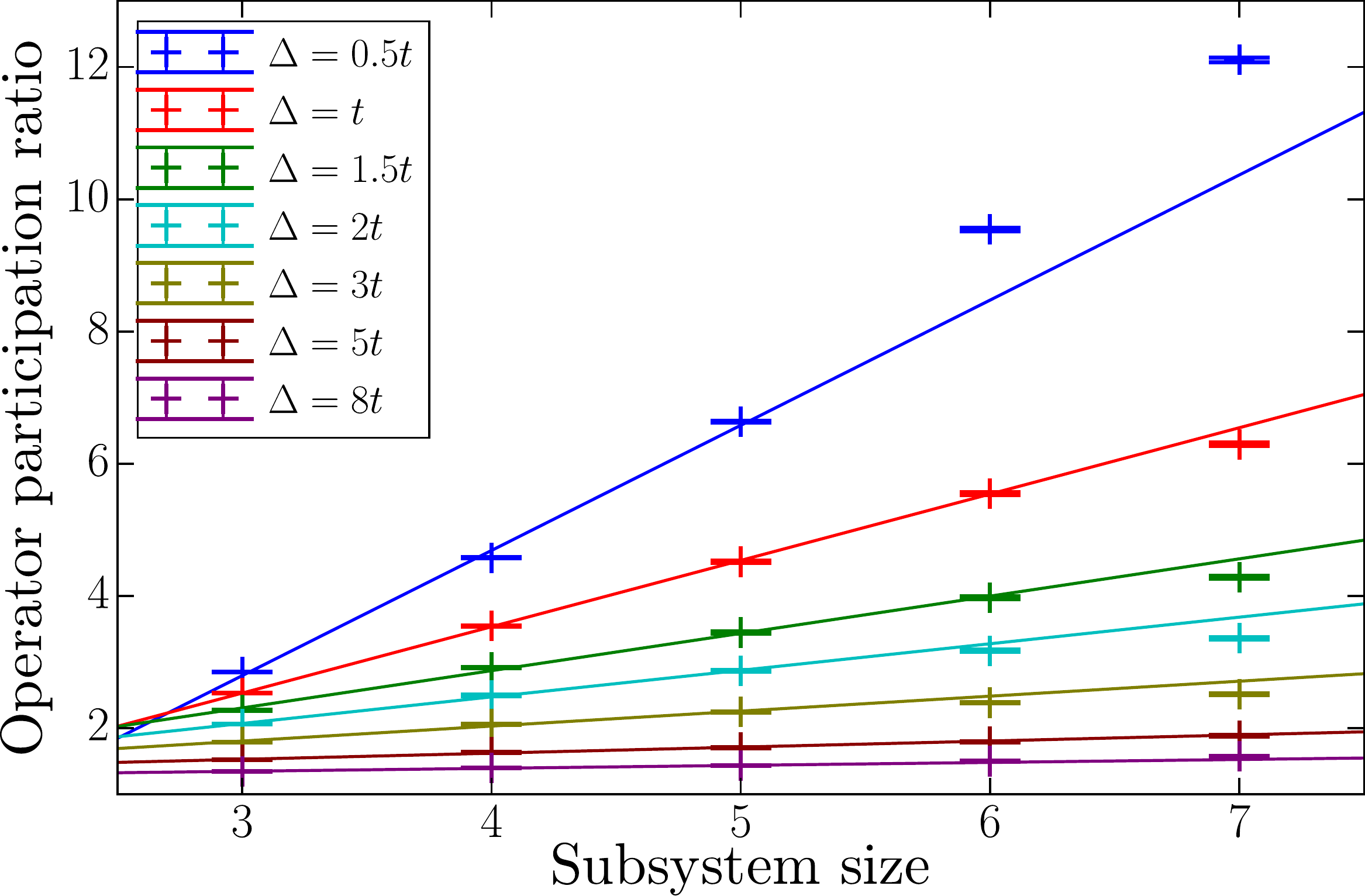}
\caption{\label{IPRFig} (Color online) Operator participation ratio, Eq. \eqref{IPRDef}, for various disorder strengths as a function of subsystem size. All points are an average over 10000 disorder realisations (error bars in plot). Lines are a fit to the first few points to emphasise trends of curving above or below a linear increase.}
\end{figure}

In Fig. \ref{IPRFig}, we plot the disorder average of $R^{-1}$ for the best-commuting operator as a function of subsystem size, for a range of disorders. We fit lines to the first few points of each curve to see the comparative trend. Deep within the localised phase, the OPR achieves a limiting value immediately and grows very little afterwards. At lower disorder, the OPR grows for a while before either bending sub-linear ($\Delta=1$), or to a higher power law ($\Delta=0.5$).

Given a local integral of motion $\taui=\sum_{l}\dtauil$, if $||\dtauil||=A\exp(-l/\xi)$, our normalisation constant is
\begin{equation}
A^2=\sum_{j=1}^\infty \exp(-2j/\xi)
\end{equation}
The value of the limit of our OPR curves ($R^{-1}(\taui)$) depends on how many Pauli operators $\Sigma_I$ from each $\dsil$ contribute to $\dtauil$, making it impossible to explicitly determine. However, it is possible to make upper and lower bounds. The minimum value of the OPR is achieved if a single operator from each $\dsil$ contributes to $\dtauil$, giving the following lower bound:
\begin{align}
R^{-1}_{\min}(\taui)&=\frac{\left[\sum_{j=0}^{\infty}\exp(-2j/\xi)\right]^2}{\sum_{j=0}^{\infty}\exp(-4j/\xi)}\nonumber\\
&=\frac{1+\exp(-2/\xi)}{1-\exp(-2/\xi)}=\coth(1/\xi),
\end{align}
whilst a bound above for the OPR can be constructed by considering the case where only the largest operator ($\sigma_i^z$) contributes:
\begin{align}
R^{-1}_{\max}(\taui)&=\left[\sum_{j=0}^{\infty}\exp(-2\xi)\right]^2=\frac{1}{(1-\exp(-2/\xi))^2}.
\end{align}
For the case of $\Delta=5t$, we have $R^{-1}(\taui)\approx 2$, which corresponds to a maximum possible decay length of $1.8$ and a minimum of $1.6$. This is over twice the decay length extracted from Fig. \ref{decay_fig}. This can be explained by the formation of singlets between pairs of LIOMs, as these would have a large effect on the OPR, but not the eigenvalue.

Whether $\lim_{l\rightarrow\infty}R^{-1}(\tauil)$ is finite is directly dependent on whether $\taui=\lim_{l\rightarrow\infty}\tauil$ is normalisable, for we can write for any $l$
\begin{equation}
\lim_{l\rightarrow\infty}R(\tauil)\geq \max(a^l_I)\frac{||\tauil||}{||\taui||} > 0,
\end{equation}
with $\max(a^l_I)$ the largest coefficient of the expansion of $\tauil$ in our Pauli basis. This is in turn related to whether the tails of $\taui$ decay faster than $1/L$ (as $1/x^n$ is a convergent series only when $n>1$). This presents a conflict; the $\Delta=0.5t$ curve is clearly diverging, whilst the corresponding eigenvalues in Fig. \ref{decay_fig} decay at a rate $1/L^{2.3}>1/L$. This is resolved by noting that the commutator $[\ham,\bracktauil]$, whilst bounded above by the size of the terms $\tauil$ on the boundary, has no bound below and can possibly scale faster than the corresponding operator. In appendix \ref{app:decay_equivalence} we demonstrate that this cannot cause a signature of exponential decay without the presence of LIOMs $\taui$.

It is unfortunately not possible to distinguish whether the sublinear behaviour of all other curves converges or is logarithmically divergent; as such we cannot determine accurately the decay behaviour close to the phase transition. It should be noted though that our method does not fix a temperature on the system. Well within the localised phase, where we approximate well the LIOMs, the operators have non-zero eigenspectrum across the entire Hilbert space, and so are at infinite temperature. However, in the presence of a mobility edge, an operator projected into the localised energy band would have better scaling than one at infinite temperature, and so this may well be observed in our method. This would imply that we expect to see signatures of the MBL phase as soon as a non-negligible fraction of the Hilbert space becomes localised. However, as a sum of a large number of operators will be required to project out delocalised energy levels, we can expect the OPR to be large (but finite) in these situations, rather than immediately collapsing to near $1$.

\begin{figure}
\includegraphics[width=\columnwidth]{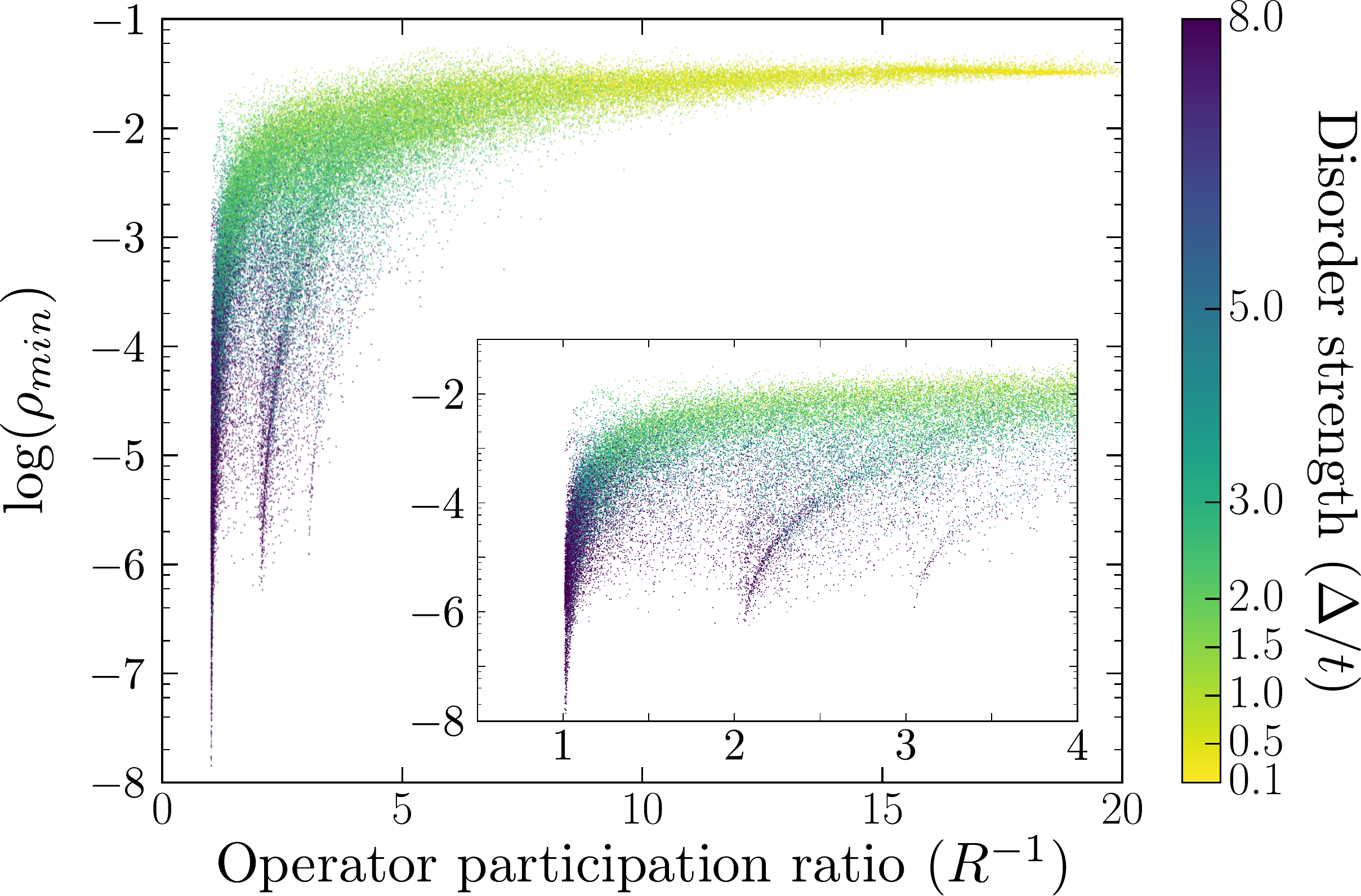}
\caption{\label{IPR_eval} (Color online) Scatter plot of $10000$ individual realisations of the lowest eigenvalue of the commutant matrix against the corresponding operator participation ratio for a range of disorders, coloured by the disorder strength $\Delta$. All points are calculated over a system size of length $7$.}
\end{figure}

\section{Correlations between operator measures}\label{sec:corr}

The model studied throughout this paper draws the disorder on any site from a uniform distribution, but this distribution is not necessarily represented well by the realisation drawn for any small subsystem we study. In the MBL phase, the density of the actual LIOMs $\taui$ within such subsystems, and the size of the tails outside, will thus fluctuate between disorder realisations. If we attempt to approximate the real LIOMs $\tau_i$ in the MBL phase with our method, we should thus expect `better' performance for some realisations over others, in the sense that the overlap between the actual LIOM $\taui$ and our ALIOM $\bracktauil$ will be larger. This implies that in the MBL phase, all measures of the ``goodness'' of our approximation should roughly correlate with each other.

We have already used the commutant matrix eigenvalues and the OPR of the resultant operators to investigate the disordered XXZ spin chain. In the MBL phase, when the eigenvalue $\rho$ corresponding to an ALIOM $\bracktauil$ is small, we expect this to be a better approximation to the LIOM $\taui$. Furthermore, we expect operators with smaller $\rho$ to have smaller tails, and thus faster decay. This should then correspond with a smaller OPR. As $R^{-1}(\taui)$ is dependent on the local structure (and thus the local disorder as well as the tails), we expect a reasonable, but not perfect correlation. 

In Fig. \ref{IPR_eval}, we make a coloured scatter plot of these two properties for a set of $10000$ disorder realisations with $L=7$, and a range of diorder strengths. We colour the plots by their disorder strength. We find a very strong correlation between the OPR and the eigenvalue; our points tend to follow a hyperboloid shape between the two limits of large and small OPR/eigenvalue. The turning point in the hyperboloid occurs at around $\Delta=2t$. 

At strong $\Delta$ and small $\rho_{min}$, we also notice a clear quantization of the OPR, with features that trend to $OPR=n$ as $\rho\rightarrow 0$ (this can be seen more clearly in the inset). These most likely correspond to the formation of singlets between two LIOMs. We note (as mentioned before) that these cause a large change to the OPR without immediately affecting the eigenvalue.

In sections \ref{sec:decay} and \ref{sec:characteristics} we made two independent measures of the decay length for the XXZ chain with parameters $\Delta=5t$, $V=0.5t$. These found $\xi=0.75$ and $\xi=1.6-1.8$ respectively, a two-fold deviation. We claim that the leading source of this deviation comes from the singlet presence, which should increase the IPR whilst not significantly affecting the eigenvalue. In order to test this claim, we remove the fraction of operators with $OPR > 2$, and recalculate both measures. The $\xi$ calculated by the eigenvalue scaling (Sec. \ref{sec:decay}) shifts to $\xi=0.72$, but the $\xi$ calculated by the OPR (Sec. \ref{sec:characteristics} shifts to $0.95-0.98$), reducing the discrepancy to $25\%$. We believe the remaining difference comes from finite size effects of the approximation.

The LIOMs $\taui$ are expected to be spin operators; they should divide the Hilbert space into two equal eigenspaces (corresponding to $\pm 1$ eigenvalues). Equivalently, being traceless, they should square to the identity. Furthermore, the results from Sec. \ref{sec:counting} suggest we should be able to isolate not only multiple $\taui$ operators, but also their products $\taui\tauj$. If this is the case, and $\taui^2=\mathbf{1}$, we can calculate $(\taui\tauj)\times\taui\times\tauj=\mathbf{1}$ also.

\begin{figure}
\begin{tabular}{cc}
\includegraphics[width=0.5\columnwidth]{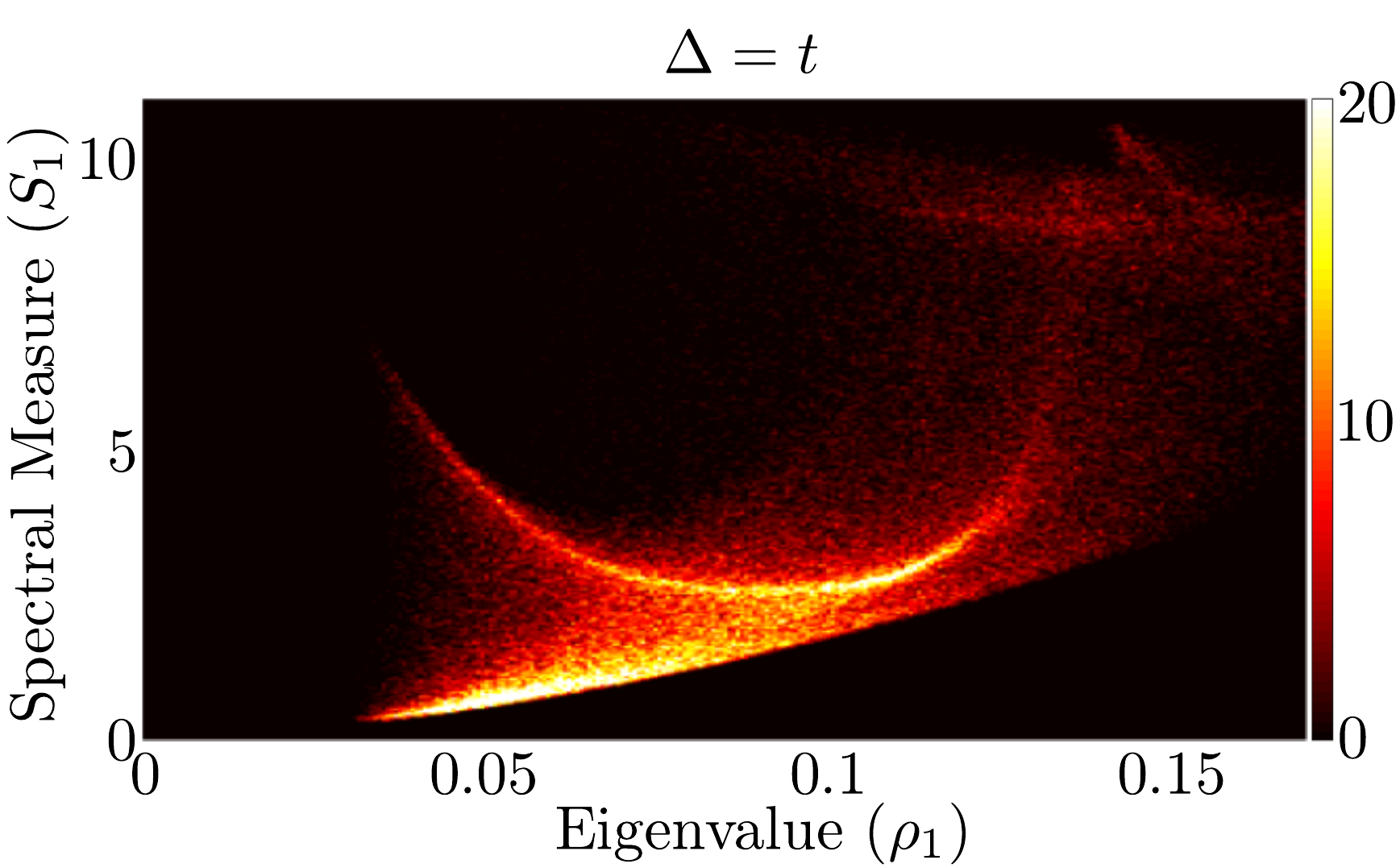}&
\includegraphics[width=0.5\columnwidth]{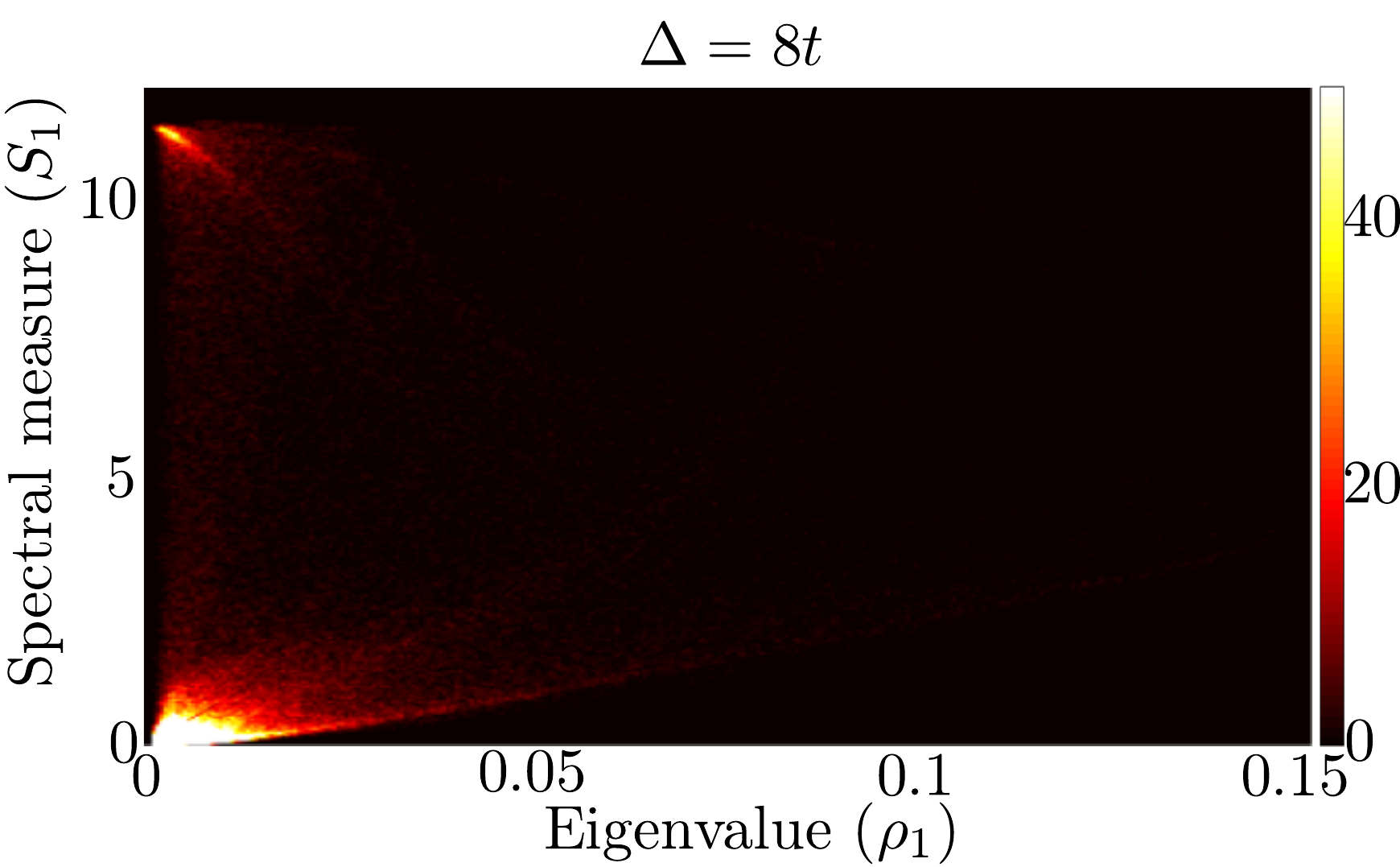}\\
\includegraphics[width=0.5\columnwidth]{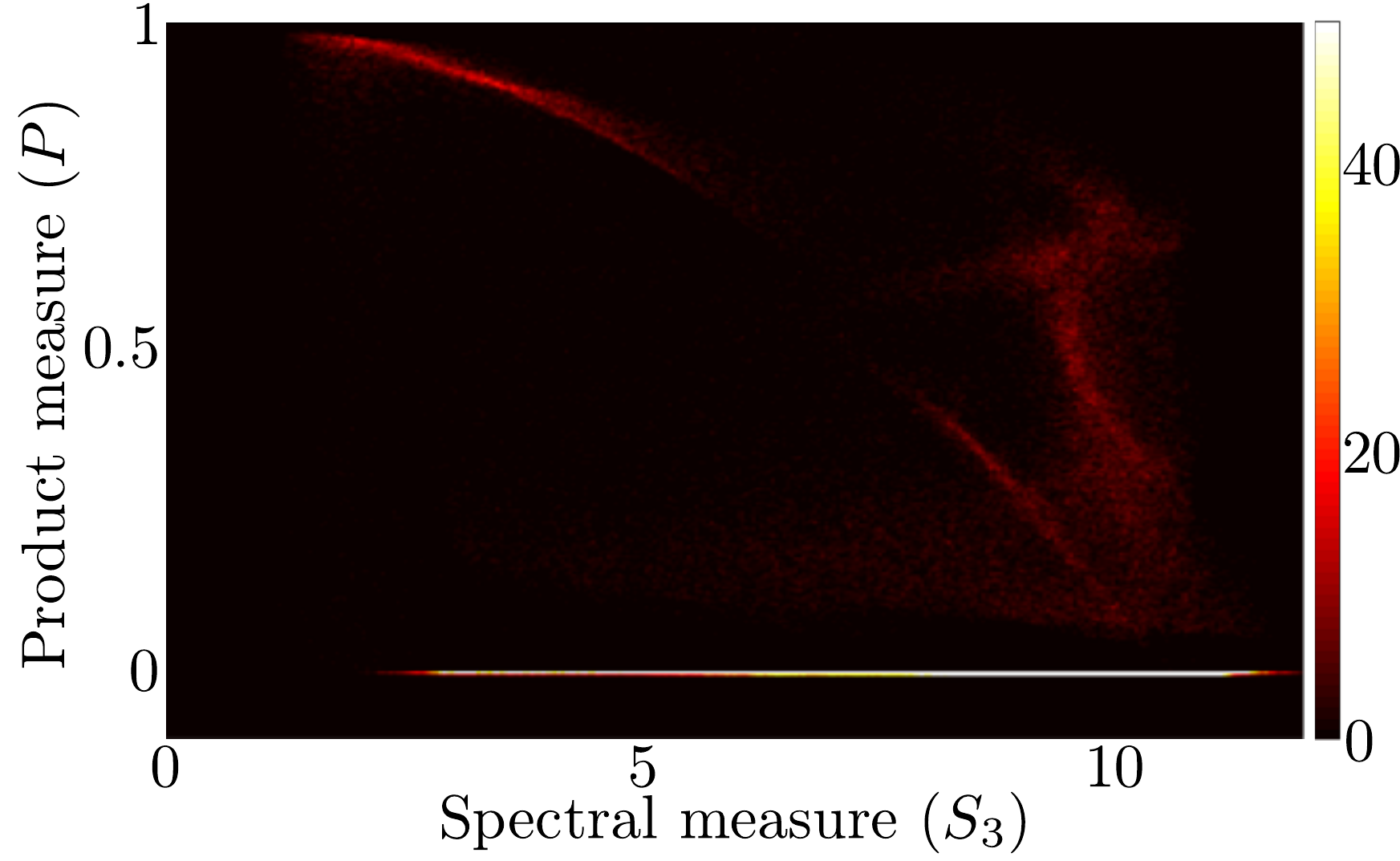}&
\includegraphics[width=0.5\columnwidth]{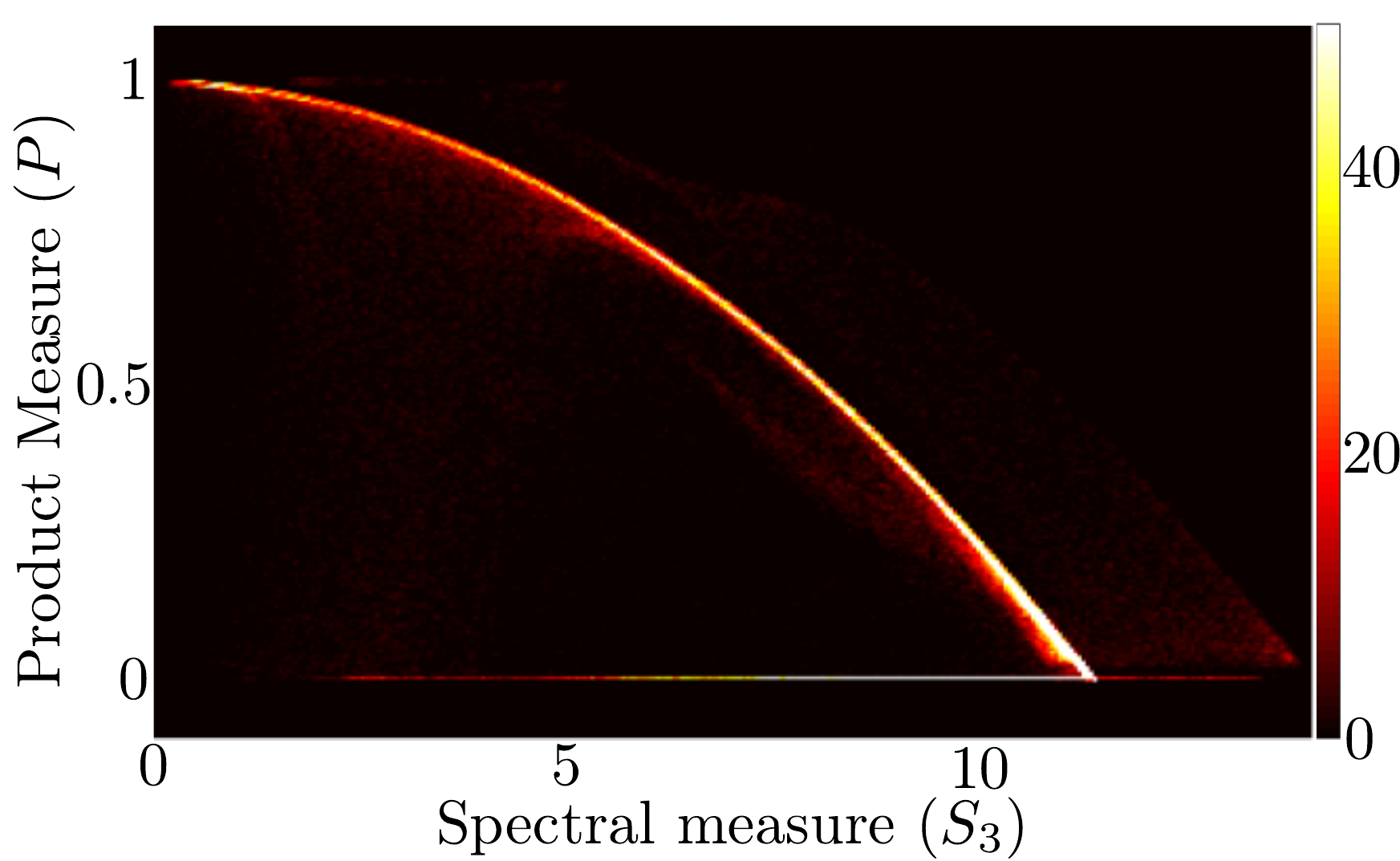}
\end{tabular}
\caption{\label{op_corr} (Color online) $2$d histogram plot of various correlations between eigenvalues and the spectral and product measures defined in text. System parameters are $V=0.5t$, and $\Delta=t$ (left) or $\Delta=8t$ (right). All data taken at a resolution of $200\times 200$ bins, from a total of $100000$ samples.}
\end{figure}

We propose two measures of the ability of the approximated $\bracktauil$ operators to mimic these algebraic properties; the spectral measure $S(\Op)$, and the product measure $P(\Op_1,\Op_2,\Op_3)$
\begin{align}
S(\Op)=\sum_{\lambda_{\Op}}||\lambda_{\Op}|-1|,
P(\Op_1,\Op_2,\Op_3)=\trace[\Op_1\Op_2\Op_3],
\end{align}
with $\lambda_{\Op}$ the eigenvalues of $\Op$. In Fig. \ref{op_corr} we plot a selection of heat maps (2d histograms) of correlations between these measures for a subsystem of size $4$ with $\Delta=t$ (left) and $\Delta=8t$ (right). The tails of operators centred at the border of the subsystem cannot exponentially decay before being cut by the approximation, so here we would expect to see a signature of two $\taui$ operators and their product. This should then produce in the $\Delta\rightarrow\infty$ limit a spectral measure $S_1=S_2=S_3=0$ for the first three eigenvalues, and $P_{123}=1$.

In the MBL phase, we see that the sample is split into two sets by the spectral measure $S_1$. The majority of the sample lies at very small $S_1$ and eigenvalue, and has a rough correlation between the two of them. A small fraction of the sample sits with spectral measure roughly $11.3$. To explain this number, we note that the Hilbert space of a four-site subsystem is $16$-dimensional, and so an operator of the form $\frac{1}{\sqrt{2}}(\sigma_i^z+\sigma_j^z)$ would have an eight-fold degenerate $0$ eigenvalue, and two four-fold degenerate $\pm \sqrt{2}$ eigenvalues. Then, we can calculate
\begin{equation}
S\left(\frac{1}{\sqrt{2}}(\sigma_i^z+\sigma_j^z)\right)= 8+8(\sqrt{2}-1)\approx 11.3.
\end{equation}
This then gives support to our explanation of the quantisation that appears in Fig. \ref{IPR_eval} (our Hilbert space is $16$-dimensional for $4$ sites, and so half of the 16 eigenvectors would have eigenvalue $0$, and the other half $\sqrt{2}$).

Furthermore, we observe a very strong correlation in the MBL phase between the spectral measure of the operator with the third-lowest eigenvalue, and the product measure, corresponding to around $90\%$ of the sample. We argue that in these samples the system is approximating two of the $\taui$ operators very well; the third is mixed somewhat with the delocalised operators, and as our approximation to the LIOM becomes better it matches simultaneously the spectral and algebraic properties. We should note here that a large fraction of the sample does have a product measure fixed to $0$ regardless of the spectral measure (about $10\%$ of the entire sample). This complete lack of overlap possibly implies that some operator other than an approximation to the LIOMs had low enough eigenvalue to sit between them.

In the ergodic phase ($\Delta=t$) there is some trend towards a lower spectral measure with lower eigenvalue. Moreover, there is an intriguing, quadratic-like feature that appears in the center of the plot, which is currently not understood. Finally, we also note that though the product measure is almost always approximately zero, there is a faint but visible remnant of the correlation that was dominant in the MBL phase.

\section{Conclusions and outlook}\label{sec:conc}

We have explicitly constructed a complete set of LIOMs, confirming the theoretical picture of the many-body localised phase of a spin system. Our method constructs the operator within a finite region that best-commutes with the Hamiltonian, which at strong enough disorder is similar to cutting off the exponentially decaying tails of the corresponding LIOM. The size of the cut tails is reflected in the norm of the commutator of this operator with the Hamiltonian, which is the eigenvalue of the constructed commutant matrix. We have observed exponential decay of the lowest eigenvalue of the commutant matrix in the MBL system (which is not present in the ergodic system), and we have proved that this is an unambiguous sign of an expontially decaying operator which commutes with the Hamiltonian. 

We have also studied the properties of these operators, quantifying their locality via an operator participation ratio, and their algebraic properties via measures of how well they approximate a set of Pauli spins. In the MBL phase, the lowest eigenvector of the commutant matrix has been shown to have the algebraic properties of a dressed spin (with the exception of a small fraction of disorder realisations where it is clearly a linear combination of a few spins). Furthermore, we have shown that this method can reproduce the entire spectrum of LIOMs living in the studied subsystems. These could alternatively be reproduced by translating the subsystem to be centered on other sites. 

The disadvantage of the method is that it scales even more prohibitively than full diagonalisation of the Hamiltonian. The commutant matrix (\ref{eq:commutant}) has size $(4^{L}-1) \times (4^{L}-1)$, making the exact method unviable for anything beyond small $L$. One could employ various schemes of truncating the basis of the operator space, either by some physical insight dependent on the particular model and/or by using MPS techniques \cite{KimCirac}. For example, a natural way of truncating the basis could be to keep only the terms that explicitly appear in the Hamiltonian and those obtained from them via the action of the commutator. This should allow to extend the method to much larger systems.

\begin{figure}[htb]
\includegraphics[width=\columnwidth]{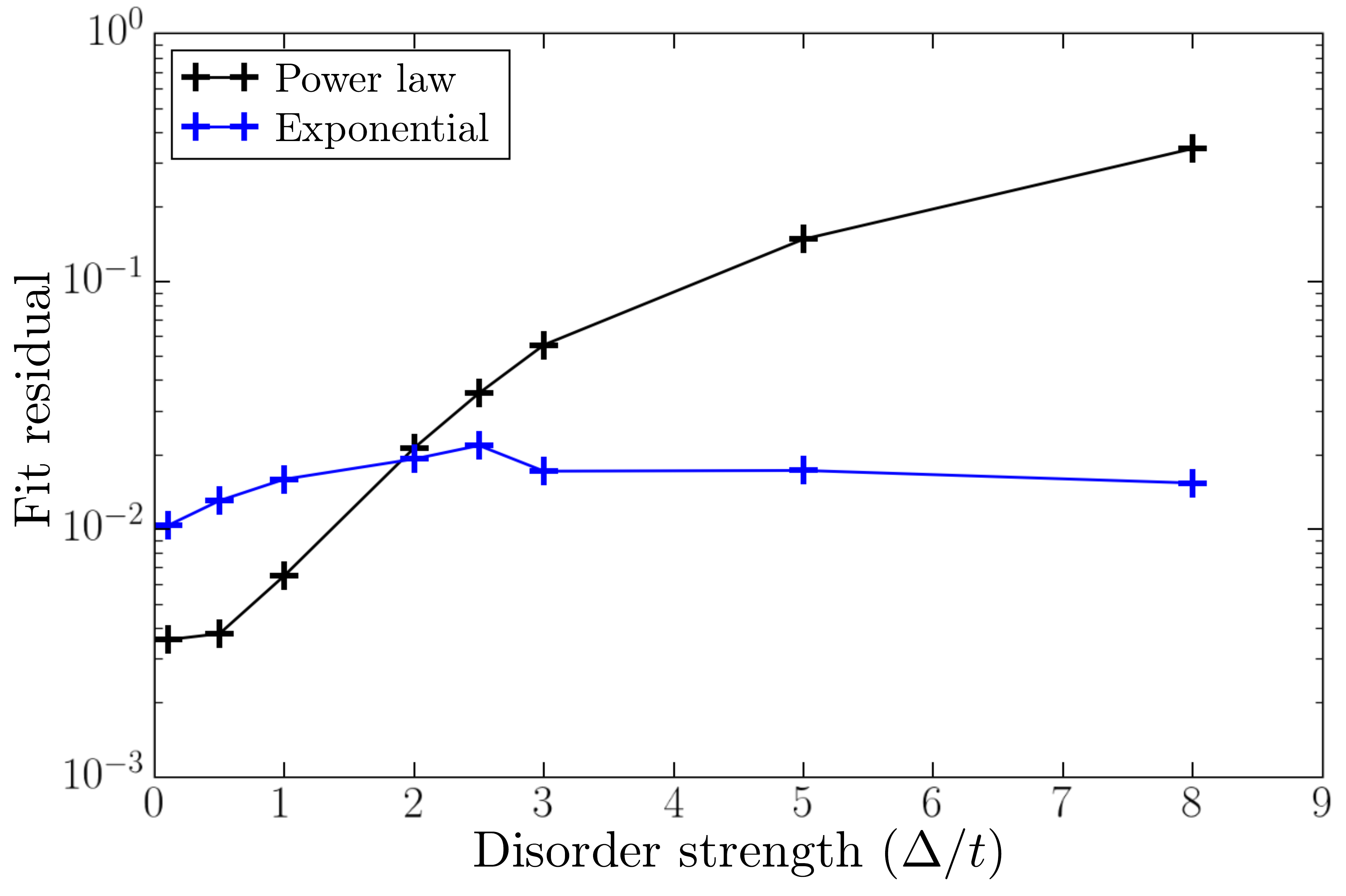}
\caption{\label{phase_transition}(Color online) Plot of the residual from the power law and ergodic fits as in Fig. \ref{decay_fig}, for a range of disorder values. Each fit was taken using $5$ data points from $L=3$ to $L=7$, each data point being in turn averaged over $1000$ disorder realisations.}
\end{figure}

Finally, a natural future direction would be to use the method presented here to study the localization-delocalization transition. The residual of the exponential and power-law fits in Sec. \ref{sec:decay} provide a means of distinguishing whether the system is in the ergodic or MBL phase. For $V=t$, we compare the two residuals in Fig. \ref{phase_transition}, and we extract a phase transition at around $\Delta=2t$. 
The critical disorder strength predicted is thus lower than in other studies, which find a mobility edge between $\Delta=2t$ and $\Delta=4t$ \cite{Serbyn15,Alet14}. This would suggest that our method ``sees" a phase transition as soon as any of the eigenspaces is localised. For this to occur, LIOMs must appear with the onset of the MBL phase in any part of the eigenspectrum. This does not appear impossible, as there exist many operators that act only on a small part of the eigenspectrum (e.g., see Ref. \onlinecite{Prosen}), and our  method would return a positive result for the MBL phase if a single one was localised. Recent work \cite{Geraedts16} appears to suggest that some LIOMs could indeed be well-defined in the presence of the mobility edge, but more detailed investigations are needed to resolve this interesting question.

\section{Acknowledgments}

We acknowledge useful discussions with Toma\v z Prosen. We also thank David Huse for several helpful comments on the paper.
T.E.O. and G.V. acknowledge support by NSERC (discovery grant).
T.E.O. also acknowledges support from the Foundation for Fundamental Research on Matter (FOM), the Netherlands Organization for Scientific Research (NWO/OCW) and an ERC Synergy Grant.
G.V. also acknowledges support by the Simons Foundation (Many Electron Collaboration). D.A. acknowledges support by SNF and by the Alfred Sloan Foundation.
Research at Perimeter Institute is
supported by the Government of Canada through Industry
Canada and by the Province of Ontario through the
Ministry of Research and Innovation.

\appendix

\section{Equivalence of two types of exponential decay}\label{app:decay_equivalence}

We wish to show that exponential decay of the lowest eigenvalue of the commutant matrix is a definite signal of localised integrals of motion. In the presence of exponentially LIOMs $\taui$, we can bound 
\begin{equation}
\minev=\frac{||[\ham,\bracktauil]||^2}{||\bracktauil||^2}\leq\frac{||[\ham,\tauil]||^2}{||\tauil||^2}\leq C'\exp(-l/\xi).
\end{equation}
This implies that our system does not have LIOMs unless $\minev$ scales exponentially. However, we have not yet demonstrated that an exponentially scaling $\minev$ must imply the presence of exponentially scaling modes. In this section, we argue why this must be the case.

There are two issues to address. Consider ``amputating" an integral of motion $\taui$ to give successive $\tauil$. We have that $[\tauil,\ham]=-[\taui-\tauil,\ham]$, and all terms in $\tauil$ have support only within $\sil$, but all terms in the remainder $\taui-\tauil$ have explicit support outside $\sil$. If $\ham$ is local, the commutator $[\ham,\taui]$ cannot have support far from that of $\taui$, but terms in $[\taui-\tauil,\ham]$ must retain support close to the outside of $\sil$, and so each term must live on the boundary. However, terms in $[\bracktauil,\ham]$ have no such restriction, as $\bracktauil$ is not constructed as part of a larger operator. This gives $\bracktauil$ additional degrees of freedom in its construction, and we must show (proposition 1) that these cannot give exponential decay that would not otherwise be present. 

Furthermore, there is no automatic lower bound for $||[\tauil,\ham]||/||\dtauil||$. We must show that this fraction cannot show exponential decay in $l$, as otherwise the commutator will see exponential decay whilst the operator itself does not. Note that this is not obviously always the case, as in the MBL phase the fraction $||[\tauil,\ham]||/||\tauil||$ \emph{does} decay exponentially. Indeed, a general proof of this has proved elusive. In proposition 2, we give a lower bound for $||[\tauil,\ham]||/||\dtauil||$ in the case of the XXZ spin chain, and point out how this cannot be immediately extended to a general model (although we hypothesise that such a lower bound always exists in non-trivial cases).

\emph{Proposition 1:} we note that the commutator between a local Hamiltonian and an operator increases linearly with the number of non-trivial tensor factors in the operator. This implies that the difference between $\tauil$ and $\bracktauil$ must be a sum of operators with small numbers of non-trivial tensor factors. The number $\numopsl$ of such operators increases polynomially in $l$, and so at large $l$, $\numopsl/\numopslp\approx1$, and our spreading operation must be independent of $l$ (i.e. it must converge in norm as a superoperator). This in turn implies that at large $l$, we should have
\begin{equation}
\frac{||[\bracktauil,\ham]||}{||[\tauil,\ham]||}\approx\frac{||[\bracktauilm,\ham]||}{||[\tauilm,\ham]||},
\end{equation}
or to be specific, the relative difference should go to zero in the $l\rightarrow\infty$ limit. Rearranging this gives
\begin{equation}
\lim_{l\rightarrow\infty}\frac{||[\bracktauil,\ham]||}{||[\bracktauilm,\ham]||}=\lim_{l\rightarrow\infty}\frac{||[\tauil,\ham]||}{||[\tauilm,\ham]||}.
\end{equation}
This implies that power law decay of $||[\tauil,\ham]||$ (right-hand side equal to $1$) is equivalent to power law decay of $||[\bracktauil,\ham]||$ (left-hand side equal to $1$).

\emph{Proposition 2:} we note that for $\taui$ to be an integral of motion, $[\tauil,\ham]$ must consist of terms from $[\dtauik,\ham]$ for $k$ near $l$. So it remains to check whether $||[\dtauil,\ham]||/||\dtauil||$ can scale exponentially. For the XXZ Hamiltonian this is not the case; an operator with non-trivial tensor factors on site $i+l$ can be written in the form
\begin{equation}
\oip = \sum_IA_I\sum_{a=x,y,z}c_{a,I}\sigma_{i+l}^a.
\end{equation}
With $A_I$ some operator acting on sites to the left of $i+l$. But then, the commutator of $\oip$ with the $\sigma_{i+l}^x\sigma_{i+l+1}^x+\sigma_{i+l}^y\sigma_{i+l+1}^y$ term gives
\begin{align}
\sum_I2A_I(c_{x,I}\sigma_{i+l}^z\sigma_{i+l+1}^y&-c_{y,I}\sigma_{i+l}^z\sigma_{i+l+1}^x\nonumber\\+&c_{z,I}\sigma_{i+l}^y\sigma_{i+l+1}^x-c_{z,I}\sigma_{i+l}^x\sigma_{i+l+1}^y),
\end{align}
giving a minimum bound 
\begin{equation}
||[\oip,\ham]||^2/||\oip||^2\geq 2
\end{equation}
The same argument applies for the operators acting on site $i-l$, and so $||[\dtauil,\ham]||/||\dtauil||$ is bounded below, and exponential decay of the numerator implies exponential decay of the denominator.

Note that Proposition 2 relied explicitly on having two terms in the Hamiltonian that acted on sites $i$ and $i+1$. In general this is not the case, and so this proof cannot be immediately extended to all systems with local Hamiltonians, although we believe the result to be true.

\bibliography{mbl}

\begin{thebibliography}{42}
\expandafter\ifx\csname natexlab\endcsname\relax\def\natexlab#1{#1}\fi
\expandafter\ifx\csname bibnamefont\endcsname\relax
  \def\bibnamefont#1{#1}\fi
\expandafter\ifx\csname bibfnamefont\endcsname\relax
  \def\bibfnamefont#1{#1}\fi
\expandafter\ifx\csname citenamefont\endcsname\relax
  \def\citenamefont#1{#1}\fi
\expandafter\ifx\csname url\endcsname\relax
  \def\url#1{\texttt{#1}}\fi
\expandafter\ifx\csname urlprefix\endcsname\relax\def\urlprefix{URL }\fi
\providecommand{\bibinfo}[2]{#2}
\providecommand{\eprint}[2][]{\url{#2}}

\bibitem[{\citenamefont{Deutsch}(1991)}]{DeutschETH}
\bibinfo{author}{\bibfnamefont{J.~M.} \bibnamefont{Deutsch}},
  \bibinfo{journal}{Phys. Rev. A} \textbf{\bibinfo{volume}{43}},
  \bibinfo{pages}{2046} (\bibinfo{year}{1991}).

\bibitem[{\citenamefont{Srednicki}(1994)}]{SrednickiETH}
\bibinfo{author}{\bibfnamefont{M.}~\bibnamefont{Srednicki}},
  \bibinfo{journal}{Phys. Rev. E} \textbf{\bibinfo{volume}{50}},
  \bibinfo{pages}{888} (\bibinfo{year}{1994}),
  \urlprefix\url{http://link.aps.org/doi/10.1103/PhysRevE.50.888}.

\bibitem[{\citenamefont{Rigol et~al.}(2008)\citenamefont{Rigol, Dunjko, and
  Olshanii}}]{RigolNature}
\bibinfo{author}{\bibfnamefont{M.}~\bibnamefont{Rigol}},
  \bibinfo{author}{\bibfnamefont{V.}~\bibnamefont{Dunjko}}, \bibnamefont{and}
  \bibinfo{author}{\bibfnamefont{M.}~\bibnamefont{Olshanii}},
  \bibinfo{journal}{Nature} \textbf{\bibinfo{volume}{452}},
  \bibinfo{pages}{854} (\bibinfo{year}{2008}).

\bibitem[{\citenamefont{Anderson}(1958)}]{Anderson58}
\bibinfo{author}{\bibfnamefont{P.~W.} \bibnamefont{Anderson}},
  \bibinfo{journal}{Phys. Rev.} \textbf{\bibinfo{volume}{109}},
  \bibinfo{pages}{1492} (\bibinfo{year}{1958}).

\bibitem[{\citenamefont{Gornyi et~al.}(2005)\citenamefont{Gornyi, Mirlin, and
  Polyakov}}]{Mirlin05}
\bibinfo{author}{\bibfnamefont{I.~V.} \bibnamefont{Gornyi}},
  \bibinfo{author}{\bibfnamefont{A.~D.} \bibnamefont{Mirlin}},
  \bibnamefont{and} \bibinfo{author}{\bibfnamefont{D.~G.}
  \bibnamefont{Polyakov}}, \bibinfo{journal}{Phys. Rev. Lett.}
  \textbf{\bibinfo{volume}{95}}, \bibinfo{pages}{206603}
  (\bibinfo{year}{2005}),
  \urlprefix\url{http://link.aps.org/doi/10.1103/PhysRevLett.95.206603}.

\bibitem[{\citenamefont{Basko et~al.}(2006)\citenamefont{Basko, Aleiner, and
  Altshuler}}]{Basko06}
\bibinfo{author}{\bibfnamefont{D.}~\bibnamefont{Basko}},
  \bibinfo{author}{\bibfnamefont{I.}~\bibnamefont{Aleiner}}, \bibnamefont{and}
  \bibinfo{author}{\bibfnamefont{B.}~\bibnamefont{Altshuler}},
  \bibinfo{journal}{Annals of Physics} \textbf{\bibinfo{volume}{321}},
  \bibinfo{pages}{1126 } (\bibinfo{year}{2006}), ISSN
  \bibinfo{issn}{0003-4916},
  \urlprefix\url{http://www.sciencedirect.com/science/article/pii/S0003491605002630}.

\bibitem[{\citenamefont{Oganesyan and Huse}(2007)}]{OganesyanHuse}
\bibinfo{author}{\bibfnamefont{V.}~\bibnamefont{Oganesyan}} \bibnamefont{and}
  \bibinfo{author}{\bibfnamefont{D.~A.} \bibnamefont{Huse}},
  \bibinfo{journal}{Phys. Rev. B} \textbf{\bibinfo{volume}{75}},
  \bibinfo{pages}{155111} (\bibinfo{year}{2007}),
  \urlprefix\url{http://link.aps.org/doi/10.1103/PhysRevB.75.155111}.

\bibitem[{\citenamefont{Znidaric et~al.}(2008)\citenamefont{Znidaric, Prosen,
  and Prelovsek}}]{Znidaric08}
\bibinfo{author}{\bibfnamefont{M.}~\bibnamefont{Znidaric}},
  \bibinfo{author}{\bibfnamefont{T.}~\bibnamefont{Prosen}}, \bibnamefont{and}
  \bibinfo{author}{\bibfnamefont{P.}~\bibnamefont{Prelovsek}},
  \bibinfo{journal}{Phys. Rev. B} \textbf{\bibinfo{volume}{77}},
  \bibinfo{pages}{064426} (\bibinfo{year}{2008}).

\bibitem[{\citenamefont{Bardarson et~al.}(2012)\citenamefont{Bardarson,
  Pollmann, and Moore}}]{Moore12}
\bibinfo{author}{\bibfnamefont{J.~H.} \bibnamefont{Bardarson}},
  \bibinfo{author}{\bibfnamefont{F.}~\bibnamefont{Pollmann}}, \bibnamefont{and}
  \bibinfo{author}{\bibfnamefont{J.~E.} \bibnamefont{Moore}},
  \bibinfo{journal}{Phys. Rev. Lett.} \textbf{\bibinfo{volume}{109}},
  \bibinfo{pages}{017202} (\bibinfo{year}{2012}),
  \urlprefix\url{http://link.aps.org/doi/10.1103/PhysRevLett.109.017202}.

\bibitem[{\citenamefont{Serbyn et~al.}(2013{\natexlab{a}})\citenamefont{Serbyn,
  Papi\ifmmode~\acute{c}\else \'{c}\fi{}, and Abanin}}]{Serbyn13-1}
\bibinfo{author}{\bibfnamefont{M.}~\bibnamefont{Serbyn}},
  \bibinfo{author}{\bibfnamefont{Z.}~\bibnamefont{Papi\ifmmode~\acute{c}\else
  \'{c}\fi{}}}, \bibnamefont{and} \bibinfo{author}{\bibfnamefont{D.~A.}
  \bibnamefont{Abanin}}, \bibinfo{journal}{Phys. Rev. Lett.}
  \textbf{\bibinfo{volume}{111}}, \bibinfo{pages}{127201}
  (\bibinfo{year}{2013}{\natexlab{a}}),
  \urlprefix\url{http://link.aps.org/doi/10.1103/PhysRevLett.111.127201}.

\bibitem[{\citenamefont{Vosk and Altman}(2013)}]{Vosk13}
\bibinfo{author}{\bibfnamefont{R.}~\bibnamefont{Vosk}} \bibnamefont{and}
  \bibinfo{author}{\bibfnamefont{E.}~\bibnamefont{Altman}},
  \bibinfo{journal}{Phys. Rev. Lett.} \textbf{\bibinfo{volume}{110}},
  \bibinfo{pages}{067204} (\bibinfo{year}{2013}).

\bibitem[{\citenamefont{Serbyn et~al.}(2014)\citenamefont{Serbyn, Knap,
  Gopalakrishnan, Papi\ifmmode~\acute{c}\else \'{c}\fi{}, Yao, Laumann, Abanin,
  Lukin, and Demler}}]{Serbyn_14_Deer}
\bibinfo{author}{\bibfnamefont{M.}~\bibnamefont{Serbyn}},
  \bibinfo{author}{\bibfnamefont{M.}~\bibnamefont{Knap}},
  \bibinfo{author}{\bibfnamefont{S.}~\bibnamefont{Gopalakrishnan}},
  \bibinfo{author}{\bibfnamefont{Z.}~\bibnamefont{Papi\ifmmode~\acute{c}\else
  \'{c}\fi{}}}, \bibinfo{author}{\bibfnamefont{N.~Y.} \bibnamefont{Yao}},
  \bibinfo{author}{\bibfnamefont{C.~R.} \bibnamefont{Laumann}},
  \bibinfo{author}{\bibfnamefont{D.~A.} \bibnamefont{Abanin}},
  \bibinfo{author}{\bibfnamefont{M.~D.} \bibnamefont{Lukin}}, \bibnamefont{and}
  \bibinfo{author}{\bibfnamefont{E.~A.} \bibnamefont{Demler}},
  \bibinfo{journal}{Phys. Rev. Lett.} \textbf{\bibinfo{volume}{113}},
  \bibinfo{pages}{147204} (\bibinfo{year}{2014}),
  \urlprefix\url{http://link.aps.org/doi/10.1103/PhysRevLett.113.147204}.

\bibitem[{\citenamefont{Huse et~al.}(2013)\citenamefont{Huse, Nandkishore,
  Oganesyan, Pal, and Sondhi}}]{HuseSondhi}
\bibinfo{author}{\bibfnamefont{D.~A.} \bibnamefont{Huse}},
  \bibinfo{author}{\bibfnamefont{R.}~\bibnamefont{Nandkishore}},
  \bibinfo{author}{\bibfnamefont{V.}~\bibnamefont{Oganesyan}},
  \bibinfo{author}{\bibfnamefont{A.}~\bibnamefont{Pal}}, \bibnamefont{and}
  \bibinfo{author}{\bibfnamefont{S.~L.} \bibnamefont{Sondhi}},
  \bibinfo{journal}{Phys. Rev. B} \textbf{\bibinfo{volume}{88}},
  \bibinfo{pages}{014206} (\bibinfo{year}{2013}),
  \urlprefix\url{http://link.aps.org/doi/10.1103/PhysRevB.88.014206}.

\bibitem[{\citenamefont{Bauer and Nayak}(2013)}]{Bauer13}
\bibinfo{author}{\bibfnamefont{B.}~\bibnamefont{Bauer}} \bibnamefont{and}
  \bibinfo{author}{\bibfnamefont{C.}~\bibnamefont{Nayak}},
  \bibinfo{journal}{Journal of Statistical Mechanics: Theory and Experiment}
  \textbf{\bibinfo{volume}{2013}}, \bibinfo{pages}{P09005}
  (\bibinfo{year}{2013}),
  \urlprefix\url{http://stacks.iop.org/1742-5468/2013/i=09/a=P09005}.

\bibitem[{\citenamefont{{Bahri} et~al.}(2013)\citenamefont{{Bahri}, {Vosk},
  {Altman}, and {Vishwanath}}}]{Bahri}
\bibinfo{author}{\bibfnamefont{Y.}~\bibnamefont{{Bahri}}},
  \bibinfo{author}{\bibfnamefont{R.}~\bibnamefont{{Vosk}}},
  \bibinfo{author}{\bibfnamefont{E.}~\bibnamefont{{Altman}}}, \bibnamefont{and}
  \bibinfo{author}{\bibfnamefont{A.}~\bibnamefont{{Vishwanath}}},
  \bibinfo{journal}{ArXiv e-prints}  (\bibinfo{year}{2013}),
  \eprint{1307.4092}.

\bibitem[{\citenamefont{Kj\"all et~al.}(2014)\citenamefont{Kj\"all, Bardarson,
  and Pollmann}}]{Kjall14}
\bibinfo{author}{\bibfnamefont{J.~A.} \bibnamefont{Kj\"all}},
  \bibinfo{author}{\bibfnamefont{J.~H.} \bibnamefont{Bardarson}},
  \bibnamefont{and} \bibinfo{author}{\bibfnamefont{F.}~\bibnamefont{Pollmann}},
  \bibinfo{journal}{Phys. Rev. Lett.} \textbf{\bibinfo{volume}{113}},
  \bibinfo{pages}{107204} (\bibinfo{year}{2014}),
  \urlprefix\url{http://link.aps.org/doi/10.1103/PhysRevLett.113.107204}.

\bibitem[{\citenamefont{Serbyn et~al.}(2013{\natexlab{b}})\citenamefont{Serbyn,
  Papi\ifmmode~\acute{c}\else \'{c}\fi{}, and Abanin}}]{Serbyn13-2}
\bibinfo{author}{\bibfnamefont{M.}~\bibnamefont{Serbyn}},
  \bibinfo{author}{\bibfnamefont{Z.}~\bibnamefont{Papi\ifmmode~\acute{c}\else
  \'{c}\fi{}}}, \bibnamefont{and} \bibinfo{author}{\bibfnamefont{D.~A.}
  \bibnamefont{Abanin}}, \bibinfo{journal}{Phys. Rev. Lett.}
  \textbf{\bibinfo{volume}{110}}, \bibinfo{pages}{260601}
  (\bibinfo{year}{2013}{\natexlab{b}}),
  \urlprefix\url{http://link.aps.org/doi/10.1103/PhysRevLett.110.260601}.

\bibitem[{\citenamefont{Huse et~al.}(2014)\citenamefont{Huse, Nandkishore, and
  Oganesyan}}]{Huse13}
\bibinfo{author}{\bibfnamefont{D.~A.} \bibnamefont{Huse}},
  \bibinfo{author}{\bibfnamefont{R.}~\bibnamefont{Nandkishore}},
  \bibnamefont{and}
  \bibinfo{author}{\bibfnamefont{V.}~\bibnamefont{Oganesyan}},
  \bibinfo{journal}{Phys. Rev. B} \textbf{\bibinfo{volume}{90}},
  \bibinfo{pages}{174202} (\bibinfo{year}{2014}),
  \urlprefix\url{http://link.aps.org/doi/10.1103/PhysRevB.90.174202}.

\bibitem[{\citenamefont{Sutherland}(2004)}]{Sutherland}
\bibinfo{author}{\bibfnamefont{B.}~\bibnamefont{Sutherland}},
  \emph{\bibinfo{title}{Beautiful Models: 70 Years of Exactly Solved Quantum
  Many-body Problems}} (\bibinfo{publisher}{World Scientific},
  \bibinfo{year}{2004}), ISBN \bibinfo{isbn}{9789812388971},
  \urlprefix\url{http://books.google.com/books?id=aVUdnwEACAAJ}.

\bibitem[{\citenamefont{Chandran et~al.}(2015)\citenamefont{Chandran, Kim,
  Vidal, and Abanin}}]{Chandran14}
\bibinfo{author}{\bibfnamefont{A.}~\bibnamefont{Chandran}},
  \bibinfo{author}{\bibfnamefont{I.~H.} \bibnamefont{Kim}},
  \bibinfo{author}{\bibfnamefont{G.}~\bibnamefont{Vidal}}, \bibnamefont{and}
  \bibinfo{author}{\bibfnamefont{D.~A.} \bibnamefont{Abanin}},
  \bibinfo{journal}{Phys. Rev. B} \textbf{\bibinfo{volume}{91}},
  \bibinfo{pages}{085425} (\bibinfo{year}{2015}),
  \urlprefix\url{http://link.aps.org/doi/10.1103/PhysRevB.91.085425}.

\bibitem[{\citenamefont{{Kim} et~al.}(2014)\citenamefont{{Kim}, {Chandran}, and
  {Abanin}}}]{Kim14}
\bibinfo{author}{\bibfnamefont{I.~H.} \bibnamefont{{Kim}}},
  \bibinfo{author}{\bibfnamefont{A.}~\bibnamefont{{Chandran}}},
  \bibnamefont{and} \bibinfo{author}{\bibfnamefont{D.~A.}
  \bibnamefont{{Abanin}}}, \bibinfo{journal}{ArXiv e-prints}
  (\bibinfo{year}{2014}), \eprint{1412.3073}.

\bibitem[{\citenamefont{Imbrie}(2014)}]{Imbrie14}
\bibinfo{author}{\bibfnamefont{J.~Z.} \bibnamefont{Imbrie}},
  \bibinfo{journal}{Arxiv e-prints}  (\bibinfo{year}{2014}),
  \eprint{1403.7837}.

\bibitem[{\citenamefont{Ros et~al.}(2015)\citenamefont{Ros, M{\"u}ller, and
  Scardicchio}}]{Ros14}
\bibinfo{author}{\bibfnamefont{V.}~\bibnamefont{Ros}},
  \bibinfo{author}{\bibfnamefont{M.}~\bibnamefont{M{\"u}ller}},
  \bibnamefont{and}
  \bibinfo{author}{\bibfnamefont{A.}~\bibnamefont{Scardicchio}},
  \bibinfo{journal}{Nuclear Physics B} \textbf{\bibinfo{volume}{891}},
  \bibinfo{pages}{420 } (\bibinfo{year}{2015}), ISSN \bibinfo{issn}{0550-3213},
  \urlprefix\url{http://www.sciencedirect.com/science/article/pii/S0550321314003836}.

\bibitem[{\citenamefont{Rademaker and Ortu\~no}(2016)}]{Rademaker}
\bibinfo{author}{\bibfnamefont{L.}~\bibnamefont{Rademaker}} \bibnamefont{and}
  \bibinfo{author}{\bibfnamefont{M.}~\bibnamefont{Ortu\~no}},
  \bibinfo{journal}{Phys. Rev. Lett.} \textbf{\bibinfo{volume}{116}},
  \bibinfo{pages}{010404} (\bibinfo{year}{2016}),
  \urlprefix\url{http://link.aps.org/doi/10.1103/PhysRevLett.116.010404}.

\bibitem[{\citenamefont{Monthus}(2016)}]{MonthusToda}
\bibinfo{author}{\bibfnamefont{C.}~\bibnamefont{Monthus}},
  \bibinfo{journal}{Journal of Physics A: Mathematical and Theoretical}
  \textbf{\bibinfo{volume}{49}}, \bibinfo{pages}{305002}
  (\bibinfo{year}{2016}),
  \urlprefix\url{http://stacks.iop.org/1751-8121/49/i=30/a=305002}.

\bibitem[{\citenamefont{He and Lu}(2016)}]{He16}
\bibinfo{author}{\bibfnamefont{R.-Q.} \bibnamefont{He}} \bibnamefont{and}
  \bibinfo{author}{\bibfnamefont{Z.-Y.} \bibnamefont{Lu}},
  \bibinfo{journal}{ArXiv e-prints}  (\bibinfo{year}{2016}),
  \eprint{1606.09509}.

\bibitem[{\citenamefont{Inglis and Pollet}(2016)}]{Inglis16}
\bibinfo{author}{\bibfnamefont{S.}~\bibnamefont{Inglis}} \bibnamefont{and}
  \bibinfo{author}{\bibfnamefont{L.}~\bibnamefont{Pollet}},
  \bibinfo{journal}{ArXiv e-prints}  (\bibinfo{year}{2016}),
  \eprint{1604.07056}.

\bibitem[{\citenamefont{Kim et~al.}(2015)\citenamefont{Kim, Ba\~nuls, Cirac,
  Hastings, and Huse}}]{KimCirac}
\bibinfo{author}{\bibfnamefont{H.}~\bibnamefont{Kim}},
  \bibinfo{author}{\bibfnamefont{M.~C.} \bibnamefont{Ba\~nuls}},
  \bibinfo{author}{\bibfnamefont{J.~I.} \bibnamefont{Cirac}},
  \bibinfo{author}{\bibfnamefont{M.~B.} \bibnamefont{Hastings}},
  \bibnamefont{and} \bibinfo{author}{\bibfnamefont{D.~A.} \bibnamefont{Huse}},
  \bibinfo{journal}{Phys. Rev. E} \textbf{\bibinfo{volume}{92}},
  \bibinfo{pages}{012128} (\bibinfo{year}{2015}),
  \urlprefix\url{http://link.aps.org/doi/10.1103/PhysRevE.92.012128}.

\bibitem[{\citenamefont{Pal and Huse}(2010)}]{PalHuse}
\bibinfo{author}{\bibfnamefont{A.}~\bibnamefont{Pal}} \bibnamefont{and}
  \bibinfo{author}{\bibfnamefont{D.~A.} \bibnamefont{Huse}},
  \bibinfo{journal}{Phys. Rev. B} \textbf{\bibinfo{volume}{82}},
  \bibinfo{pages}{174411} (\bibinfo{year}{2010}),
  \urlprefix\url{http://link.aps.org/doi/10.1103/PhysRevB.82.174411}.

\bibitem[{\citenamefont{Kim et~al.}(2014)\citenamefont{Kim, Ikeda, and
  Huse}}]{Kim_ETH}
\bibinfo{author}{\bibfnamefont{H.}~\bibnamefont{Kim}},
  \bibinfo{author}{\bibfnamefont{T.~N.} \bibnamefont{Ikeda}}, \bibnamefont{and}
  \bibinfo{author}{\bibfnamefont{D.~A.} \bibnamefont{Huse}},
  \bibinfo{journal}{Phys. Rev. E} \textbf{\bibinfo{volume}{90}},
  \bibinfo{pages}{052105} (\bibinfo{year}{2014}),
  \urlprefix\url{http://link.aps.org/doi/10.1103/PhysRevE.90.052105}.

\bibitem[{\citenamefont{Kim and Huse}(2013)}]{KimBallistic}
\bibinfo{author}{\bibfnamefont{H.}~\bibnamefont{Kim}} \bibnamefont{and}
  \bibinfo{author}{\bibfnamefont{D.~A.} \bibnamefont{Huse}},
  \bibinfo{journal}{Phys. Rev. Lett.} \textbf{\bibinfo{volume}{111}},
  \bibinfo{pages}{127205} (\bibinfo{year}{2013}),
  \urlprefix\url{http://link.aps.org/doi/10.1103/PhysRevLett.111.127205}.

\bibitem[{\citenamefont{\ifmmode \check{Z}\else
  \v{Z}\fi{}nidari\ifmmode~\check{c}\else \v{c}\fi{}
  et~al.}(2016)\citenamefont{\ifmmode \check{Z}\else
  \v{Z}\fi{}nidari\ifmmode~\check{c}\else \v{c}\fi{}, Scardicchio, and
  Varma}}]{Znidaric16}
\bibinfo{author}{\bibfnamefont{M.}~\bibnamefont{\ifmmode \check{Z}\else
  \v{Z}\fi{}nidari\ifmmode~\check{c}\else \v{c}\fi{}}},
  \bibinfo{author}{\bibfnamefont{A.}~\bibnamefont{Scardicchio}},
  \bibnamefont{and} \bibinfo{author}{\bibfnamefont{V.~K.} \bibnamefont{Varma}},
  \bibinfo{journal}{Phys. Rev. Lett.} \textbf{\bibinfo{volume}{117}},
  \bibinfo{pages}{040601} (\bibinfo{year}{2016}),
  \urlprefix\url{http://link.aps.org/doi/10.1103/PhysRevLett.117.040601}.

\bibitem[{\citenamefont{Luitz et~al.}(2015)\citenamefont{Luitz, Laflorencie,
  and Alet}}]{Alet14}
\bibinfo{author}{\bibfnamefont{D.~J.} \bibnamefont{Luitz}},
  \bibinfo{author}{\bibfnamefont{N.}~\bibnamefont{Laflorencie}},
  \bibnamefont{and} \bibinfo{author}{\bibfnamefont{F.}~\bibnamefont{Alet}},
  \bibinfo{journal}{Phys. Rev. B} \textbf{\bibinfo{volume}{91}},
  \bibinfo{pages}{081103} (\bibinfo{year}{2015}),
  \urlprefix\url{http://link.aps.org/doi/10.1103/PhysRevB.91.081103}.

\bibitem[{\citenamefont{Serbyn et~al.}(2015)\citenamefont{Serbyn,
  Papi\ifmmode~\acute{c}\else \'{c}\fi{}, and Abanin}}]{Serbyn15}
\bibinfo{author}{\bibfnamefont{M.}~\bibnamefont{Serbyn}},
  \bibinfo{author}{\bibfnamefont{Z.}~\bibnamefont{Papi\ifmmode~\acute{c}\else
  \'{c}\fi{}}}, \bibnamefont{and} \bibinfo{author}{\bibfnamefont{D.~A.}
  \bibnamefont{Abanin}}, \bibinfo{journal}{Phys. Rev. X}
  \textbf{\bibinfo{volume}{5}}, \bibinfo{pages}{041047} (\bibinfo{year}{2015}),
  \urlprefix\url{http://link.aps.org/doi/10.1103/PhysRevX.5.041047}.

\bibitem[{\citenamefont{De~Roeck et~al.}(2016)\citenamefont{De~Roeck,
  Huveneers, M\"uller, and Schiulaz}}]{SchiulazMobility}
\bibinfo{author}{\bibfnamefont{W.}~\bibnamefont{De~Roeck}},
  \bibinfo{author}{\bibfnamefont{F.}~\bibnamefont{Huveneers}},
  \bibinfo{author}{\bibfnamefont{M.}~\bibnamefont{M\"uller}}, \bibnamefont{and}
  \bibinfo{author}{\bibfnamefont{M.}~\bibnamefont{Schiulaz}},
  \bibinfo{journal}{Phys. Rev. B} \textbf{\bibinfo{volume}{93}},
  \bibinfo{pages}{014203} (\bibinfo{year}{2016}),
  \urlprefix\url{http://link.aps.org/doi/10.1103/PhysRevB.93.014203}.

\bibitem[{\citenamefont{Friesdorf et~al.}(2015)\citenamefont{Friesdorf, Werner,
  Brown, Scholz, and Eisert}}]{Friesdorf}
\bibinfo{author}{\bibfnamefont{M.}~\bibnamefont{Friesdorf}},
  \bibinfo{author}{\bibfnamefont{A.~H.} \bibnamefont{Werner}},
  \bibinfo{author}{\bibfnamefont{W.}~\bibnamefont{Brown}},
  \bibinfo{author}{\bibfnamefont{V.~B.} \bibnamefont{Scholz}},
  \bibnamefont{and} \bibinfo{author}{\bibfnamefont{J.}~\bibnamefont{Eisert}},
  \bibinfo{journal}{Phys. Rev. Lett.} \textbf{\bibinfo{volume}{114}},
  \bibinfo{pages}{170505} (\bibinfo{year}{2015}),
  \urlprefix\url{http://link.aps.org/doi/10.1103/PhysRevLett.114.170505}.

\bibitem[{\citenamefont{Eisert et~al.}(2010)\citenamefont{Eisert, Cramer, and
  Plenio}}]{Plenio}
\bibinfo{author}{\bibfnamefont{J.}~\bibnamefont{Eisert}},
  \bibinfo{author}{\bibfnamefont{M.}~\bibnamefont{Cramer}}, \bibnamefont{and}
  \bibinfo{author}{\bibfnamefont{M.~B.} \bibnamefont{Plenio}},
  \bibinfo{journal}{Rev. Mod. Phys.} \textbf{\bibinfo{volume}{82}},
  \bibinfo{pages}{277} (\bibinfo{year}{2010}),
  \urlprefix\url{http://link.aps.org/doi/10.1103/RevModPhys.82.277}.

\bibitem[{\citenamefont{Prosen}(2011)}]{Prosen2011}
\bibinfo{author}{\bibfnamefont{T.}~\bibnamefont{Prosen}},
  \bibinfo{journal}{Phys. Rev. Lett.} \textbf{\bibinfo{volume}{106}},
  \bibinfo{pages}{217206} (\bibinfo{year}{2011}),
  \urlprefix\url{http://link.aps.org/doi/10.1103/PhysRevLett.106.217206}.

\bibitem[{\citenamefont{Pereira et~al.}(2014)\citenamefont{Pereira, Pasquier,
  Sirker, and Affleck}}]{Pereira2014}
\bibinfo{author}{\bibfnamefont{R.~G.} \bibnamefont{Pereira}},
  \bibinfo{author}{\bibfnamefont{V.}~\bibnamefont{Pasquier}},
  \bibinfo{author}{\bibfnamefont{J.}~\bibnamefont{Sirker}}, \bibnamefont{and}
  \bibinfo{author}{\bibfnamefont{I.}~\bibnamefont{Affleck}},
  \bibinfo{journal}{Journal of Statistical Mechanics: Theory and Experiment}
  \textbf{\bibinfo{volume}{2014}}, \bibinfo{pages}{P09037}
  (\bibinfo{year}{2014}),
  \urlprefix\url{http://stacks.iop.org/1742-5468/2014/i=9/a=P09037}.

\bibitem[{\citenamefont{Ilievski et~al.}(2015)\citenamefont{Ilievski, Medenjak,
  and Prosen}}]{Ilievski2015}
\bibinfo{author}{\bibfnamefont{E.}~\bibnamefont{Ilievski}},
  \bibinfo{author}{\bibfnamefont{M.}~\bibnamefont{Medenjak}}, \bibnamefont{and}
  \bibinfo{author}{\bibfnamefont{T.}~\bibnamefont{Prosen}},
  \bibinfo{journal}{Phys. Rev. Lett.} \textbf{\bibinfo{volume}{115}},
  \bibinfo{pages}{120601} (\bibinfo{year}{2015}),
  \urlprefix\url{http://link.aps.org/doi/10.1103/PhysRevLett.115.120601}.

\bibitem[{\citenamefont{Mierzejewski et~al.}(2015)\citenamefont{Mierzejewski,
  Prosen, and Prelovsek}}]{Prosen}
\bibinfo{author}{\bibfnamefont{M.}~\bibnamefont{Mierzejewski}},
  \bibinfo{author}{\bibfnamefont{T.}~\bibnamefont{Prosen}}, \bibnamefont{and}
  \bibinfo{author}{\bibfnamefont{P.}~\bibnamefont{Prelovsek}},
  \bibinfo{journal}{Phys. Rev. B} \textbf{\bibinfo{volume}{92}},
  \bibinfo{pages}{195121} (\bibinfo{year}{2015}),
  \urlprefix\url{http://link.aps.org/doi/10.1103/PhysRevB.92.195121}.

\bibitem[{\citenamefont{Geraedts et~al.}(2016)\citenamefont{Geraedts, Bhatt,
  and Nandkishore}}]{Geraedts16}
\bibinfo{author}{\bibfnamefont{S.~D.} \bibnamefont{Geraedts}},
  \bibinfo{author}{\bibfnamefont{R.}~\bibnamefont{Bhatt}}, \bibnamefont{and}
  \bibinfo{author}{\bibfnamefont{R.}~\bibnamefont{Nandkishore}},
  \bibinfo{journal}{ArXiv e-prints}  (\bibinfo{year}{2016}),
  \eprint{1608.01328}.

\end{thebibliography}

\end{document}